\documentclass[sn-nature]{sn-jnl}

\usepackage{graphicx}%
\graphicspath{{./images/}}
\usepackage{multirow}%
\usepackage{amsmath,amssymb,amsfonts}%
\usepackage{amsthm}%
\usepackage{mathrsfs}%
\usepackage[title]{appendix}%
\usepackage{xcolor}%
\usepackage{textcomp}%
\usepackage{manyfoot}%
\usepackage{algorithm}%
\usepackage{algorithmicx}%
\usepackage{algpseudocode}%
\usepackage{listings}%
\usepackage{soul}%
\usepackage{booktabs,makecell,pifont}
\makeatletter
\let\cline\cmidrule
\makeatother
\usepackage{pdfcomment}
\usepackage{longtable}
\usepackage{float}
\usepackage{hyperref}
\usepackage{geometry}
\makeatletter
\long\def\@figurecaption#1#2{{\figurecaptionfont{\bfseries#1\ \textbar\ }#2\par}}%
\makeatother

\renewcommand{\thefigure}{\arabic{figure}} 
\renewcommand{\thetable}{\arabic{table}}

\theoremstyle{thmstyleone}%
\theoremstyle{thmstyletwo}%

\theoremstyle{thmstylethree}%

\begin{document}

\title{Scalable Prediction of Complex Surface Reconstructions under Operating Conditions via Harmony-Search-Based Global Optimization}

\author[1]{\fnm{Shiyang} \sur{Chen}}

\author*[1,2]{\fnm{Xiao-Ming} \sur{Cao}}\email{xmcao@sjtu.edu.cn}

\affil*[1]{\orgdiv{State Key Laboratory of Green Chemical Engineering and Industrial Catalysis, Centre for Computational Chemistry and Research Institute of Industrial Catalysis}, \orgname{East China University of Science and Technology}, \orgaddress{\street{130 Meilong Road}, \city{Shanghai}, \postcode{200237}, \state{Shanghai}, \country{China}}}

\affil*[2]{\orgdiv{State Key Laboratory of Synergistic Chem-Bio Synthesis, School of Chemistry and Chemical Engineering}, \orgname{Shanghai Jiao Tong University}, \orgaddress{\street{800 Dongchuan Road}, \city{Shanghai}, \postcode{200240}, \state{Shanghai}, \country{China}}}

\abstract{The dynamic structural evolution of catalyst surfaces under operating conditions dictates catalytic performance, yet capturing these reconstructions atomically remains challenging. Global optimization based on machine learning interatomic potentials (MLIPs) is promising, but scaling to large-scale, low-symmetry \textit{operando} systems is hindered by expansive search spaces and potential energy surface (PES) inaccuracies. Herein, we present Harmony-search-based Atomic Structural Global Optimization (HASGO), a framework integrating universal MLIPs with a harmony search algorithm. HASGO overcomes the problem of PES softening by incorporating a multi-head replay fine-tuning protocol. Moreover, the stochastic structural perturbation step in its algorithm offers a fault-tolerant strategy to enhance the robustness of global convergence. These enable HASGO to identify intricate surface oxide overlayers that align with atomic-resolution microscopy, thereby resolving the square-pyramidal subsurface O$_5$ motif on Ag(100) during ethylene epoxidation. This scalable framework provides a robust approach for uncovering \textit{operando} structures, accelerating the rational design of industrial catalysts.}

\maketitle

\section*{Introduction}
Transition metals exhibit exceptional catalytic activity in a wide range of reactions across chemical production, environmental treatment, and energy conversion. The interplay between metal and reaction intermediates frequently induces surface reconstructions and phase transitions, causing surface active sites to evolve dynamically during reaction\cite{MetalOxide,Surfacereconstructions}. For instance, dynamically generated surface oxide species serve as the realistic active centres in many oxidation processes, such as CO oxidation\cite{PdOnaturegas1,PdOnaturegas2}, methane combustion\cite{PdOmethane1,PdOmethane2,CuOmethane,Pd-Pt_methane}, and the oxygen evolution reaction\cite{H2Ooxidation1,H2Ooxidation2}. Nevertheless, elucidating the evolution of surface structures at the atomic level under the high-temperature and high-pressure conditions typical of industrial catalysis remains a significant challenge for current \textit{operando} characterisation techniques.

Experimentally observed surface reconstructions at low pressure often correspond to the most thermodynamically stable structures, representing the global minimum (GM) on the PES\cite{agox,GCGO_hammer,HEA,ASOPs,lzp_Ag4OAg}. This alignment suggests that surface reconstruction can be rigorously modelled as a global optimisation (GO) problem. While the advancement of first-principles calculations has enabled the application of GO algorithms, such as USPEX\cite{USPEX}, CALYPSO\cite{CALYPSO}, and ABCluster\cite{ABCluster}, to simulate these surfaces, substantial computational costs have largely restricted research to small unit cells, typically \textit{p}(2$\times$2)\cite{Limitsize1,Limitsize2,Limitsize3}. Importantly, surface reconstruction under operating conditions is a process governed by the Grand Canonical (GC) ensemble, characterised by fixed chemical potentials but variable stoichiometry and symmetry. As recently demonstrated by Liu and co-workers\cite{ASOPs}, confining simulations to small, high-symmetry unit cells severely constrains the searchable chemical space, potentially precluding the discovery of the true thermodynamically stable phases under operating conditions. Consequently, there is a clear imperative for techniques that can mitigate the computational burden of GO on large-scale, low-symmetry systems while maintaining first-principles accuracy.

Machine learning interatomic potentials (MLIPs) have emerged as a promising alternative to first-principles calculations for accelerating PES exploration. Equivariant graph neural network (GNN) architectures, such as PAINN\cite{Painn}, MACE\cite{MACE}, Orb\cite{ORB}, and Equiformer\cite{Equiformer}, have significantly reduced the data requirements for training while maintaining near-density functional theory (DFT) accuracy\cite{ASOPs,lzp_Ag4OAg,PES_explore_GCMC,PES_explore_MD,PES_explore_BH,PES_explore_dock,EMLP,AL-PES_explore}. However, applying MLIP-based GO to large surfaces presents two primary challenges. First, the limited extrapolation capabilities of MLIPs necessitate that investigated configurations remain within, or in proximity to, the structural distribution of the training set, requiring extensive DFT labelling. Second, as system size increases, the high-dimensional nature of the PES demands a sophisticated balance between exploration and exploitation to avoid entrapment in spurious local minima due to MLIP inaccuracies.

Recent advances in universal MLIPs (u-MLIPs) have demonstrated impressive ``out-of-the-box'' modelling capabilities across the entire periodic table\cite{chgnetfoundation,unifiedfoundation,chen2022universal,merchant2023scaling,zhang2024pretraining,mattersim,foundation-long_range,u-MLIP_application}. While u-MLIPs can reduce initial training costs, they frequently suffer from systematic ``PES softening,'' where atomic forces and energies are systematically underestimated\cite{finetune-soften,chen2022universal,merchant2023scaling}. This softening is often rooted in the narrow distribution of pre-training data, which lacks the high-energy, non-equilibrium configurations encountered during GM search\cite{non-equilibrium_data,huosongze}. Although fine-tuning with out-of-distribution (OOD) samples can enhance robustness, it often triggers ``catastrophic forgetting'', eroding the model's broad chemical knowledge\cite{catastrophic_forgetting_1,catastrophic_forgetting_2,catastrophic_forgetting_3}. Furthermore, a robust GO framework must be inherently fault-tolerant, capable of accommodating marginal PES deviations between the MLIP and DFT benchmarks, especially during the exploration of high-energy regions. Thus, achieving synergy between data strategies and GO algorithms is a key factor for the efficient search and identification of realistic surface structures under operating conditions.

Inspired by these advances, we developed Harmony-search-based Atomic Structural Global Optimization (HASGO), a framework that integrates system-specific MLIPs with the harmony search (HS) algorithm\cite{HS} for efficient atomistic GO under realistic catalytic environments. In HASGO, the MLIP is constructed by fine-tuning a u-MLIP with DFT data generated during GO on a smaller surrogate system. By leveraging the HS algorithm's ability to strike a balance between exploration and exploitation, HASGO accelerates the structural search while simultaneously building a reliable and data-efficient dataset for model refinement. To evaluate its performance in realistic surface reconstruction problems, we applied variable-lattice HASGO to metal oxide reconstruction on Pd(100), Cu(100), and Ag(100). Variable-lattice HASGO first reproduces experimentally observed high-vacuum surface oxide phases, including the PdO(101)-step reconstruction on \textit{p}(5$\times$5) Pd(100), the $(2\sqrt{2}\times\sqrt{2})$R45$^\circ$ missing-row reconstruction (MRR) CuO phase on Cu(100), and the analogous MRR AgO phase on Ag(100). We further show that, under the operating condition of ethylene epoxidation, variable-lattice HASGO identifies an Ag(100)-O reconstruction featuring a subsurface O$_5$ motif, in line with recent infrared spectroscopic discovery\cite{lzp_Ag4OAg}. These validations demonstrate HASGO's broad applicability across diverse chemical environments and structural scales, providing a scalable pathway for uncovering the \textit{operando} structures critical to industrial catalysis.

\section*{Results}

\subsection*{An Overview of the HASGO Workflow}
\begin{figure}[H]
  \centering
  \includegraphics[scale=0.3]{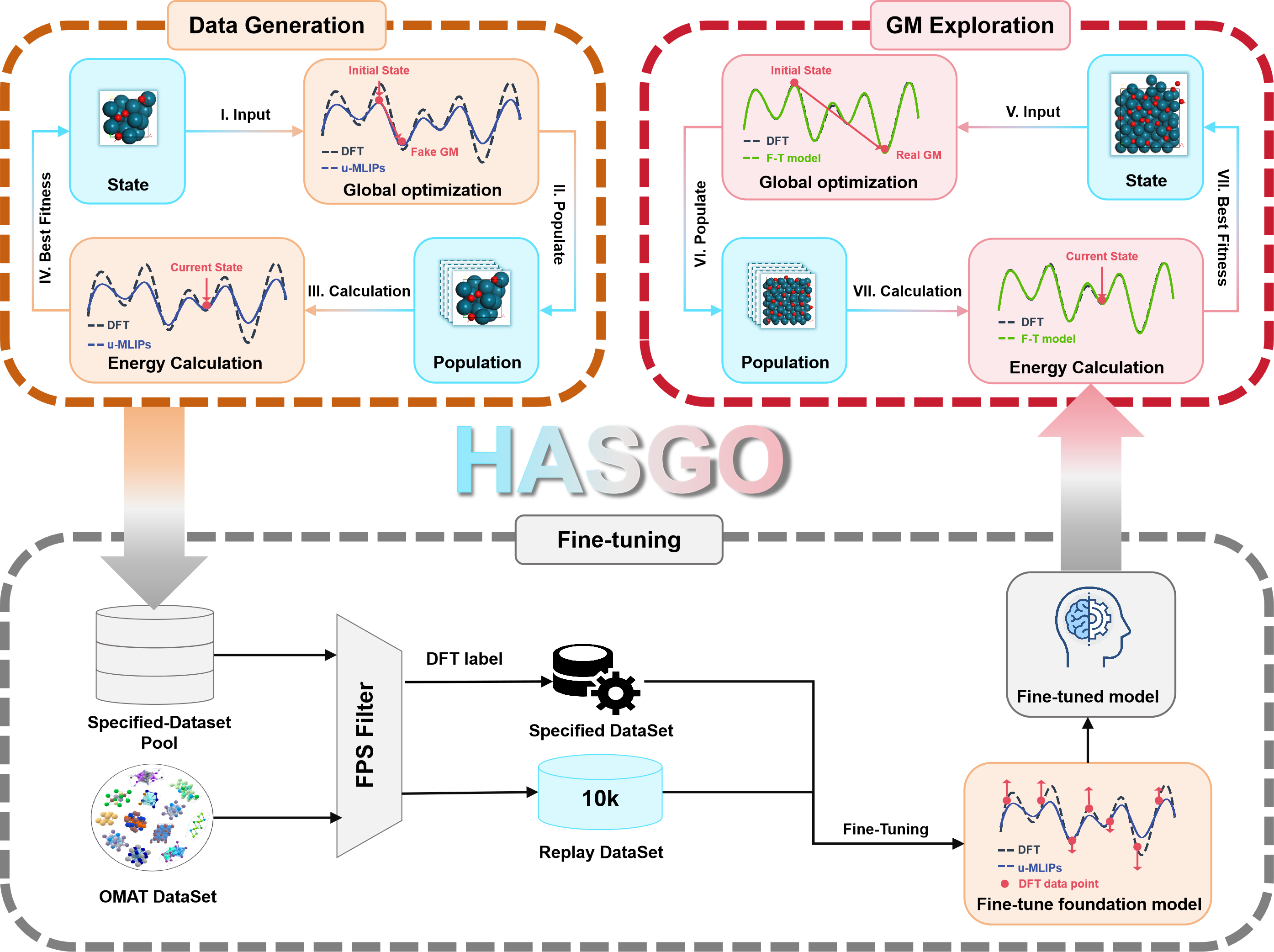}
  \caption{\textbf{Overview of the HASGO framework}. The workflow consists of the data generation part (upper left), the fine-tuning part (lower), and the GM exploration part (upper right). Data generation: (I) initial GO using HS algorithms for target systems with limited size, (II) GO using the foundation model for (III) energy calculation and fitness evaluation to (IV) identify the small-scale GM structure. Fine-tuning: local optimisation trajectories of small-scale target systems are sampled by the FPS strategy and curated into a specified dataset after labelling, which is used to fine-tune the foundation model via a multi-head replay training strategy for GM exploration. GM exploration: (V) initial GO of larger-scale target systems followed by (VI) enhanced GO using the fine-tuned model for (VII) accurate energy calculation and fitness evaluation, and (VIII) identification of the best fitness structure with the lowest potential energy as the larger-scale GM.}
  \label{fig:hasgo}
\end{figure}

Fig.~\ref{fig:hasgo} shows the workflow of HASGO. It integrates GO via the HS algorithm with the fine-tuning of a u-MLIP to efficiently navigate the reliable PES of target systems. Dataset generation for u-MLIP fine-tuning proceeds in three stages. Initially, GO utilising the HS algorithm and the foundation u-MLIP generates candidate structures for small-scale systems (typically $\leq$ 25~atoms). Subsequently, a representative subset is extracted using the farthest point sampling (FPS) strategy\cite{FPS}, drawing from all generated candidates and local optimisation trajectories. These selected structures are labelled via DFT single-point calculations and used alongside a multi-head replay training protocol\cite{multi-head-replay} to adapt the original u-MLIP to the specific target system.

During the primary GM exploration phase, this fine-tuned u-MLIP enables the metaheuristic algorithm to reliably navigate the solution space. It provides rapid fitness evaluations and local relaxations in the downstream GO cycle, ultimately identifying the configuration with the lowest energy. This workflow mitigates the limited extrapolation capabilities of MLIPs when encountering OOD structures, ensuring robust energy evaluations throughout the optimisation process. In addition, the GM exploration at each epoch undergoes a random choice between exploration and exploitation, followed by structural perturbation in the HS algorithm. This structural perturbation step provides fault tolerance against marginal discrepancies between the MLIP and DFT PES, enhancing the robustness of GM convergence. Detailed HASGO configurations are provided in the Methods section.

\subsection*{Harmony Search Effectiveness}

To identify the optimal GO algorithm for locating the GM on the PES, we conducted a systematic benchmark of ten prominent GO algorithms. The evaluation was performed, using a fixed 500 epochs for GO with the MACE-OMAT foundation model for local relaxation and energy calculation, on a benchmark $p(4 \times 4)$ Pd(100) surface model containing oxygen vacancies\cite{HEA}, a larger system of critical importance for controlling catalytic reaction rates\cite{oxygen_vacancy_1,oxygen_vacancy_2}. The benchmark algorithms included Genetic Algorithm (GA)\cite{GA}, Particle Swarm Optimization (PSO)\cite{PSO}, Ant Colony Optimization (ACO)\cite{ACO}, Artificial Bee Colony (ABC)\cite{ABCluster}, Grey Wolf Optimizer (GWO)\cite{GWO}, Improved GWO (IGWO)\cite{IGWO}, Whale Optimization Algorithm (WOA)\cite{WOA}, Biogeography-Based Optimizer (BBO)\cite{BBO}, Multi-Verse Optimizer (MVO)\cite{MVO}, and HS\cite{HS}.

Our results demonstrate that the HS algorithm exhibits the best balance between computational efficiency and search reliability. As illustrated in the comparative analysis (Fig.~\ref{fig:finetune-energy-comparison}a and Fig.~\ref{fig:finetune-energy-comparison}b), HS identified a GM configuration significantly more stable than those found by competing algorithms. This robust performance persists even as population sizes scale (Fig.~\ref{fig:finetune-energy-comparison}b), establishing HS as the effective metaheuristic for high-dimensional atomic structural searches. Furthermore, with the increase of population sizes, HS achieves more efficient convergence while exhibiting a consistent GM with that obtained using a small population size.

\subsection*{Energy Validation of the Fine-Tuned u-MLIP Model}
\begin{figure}[H]
  \centering
  \includegraphics[scale=0.25]{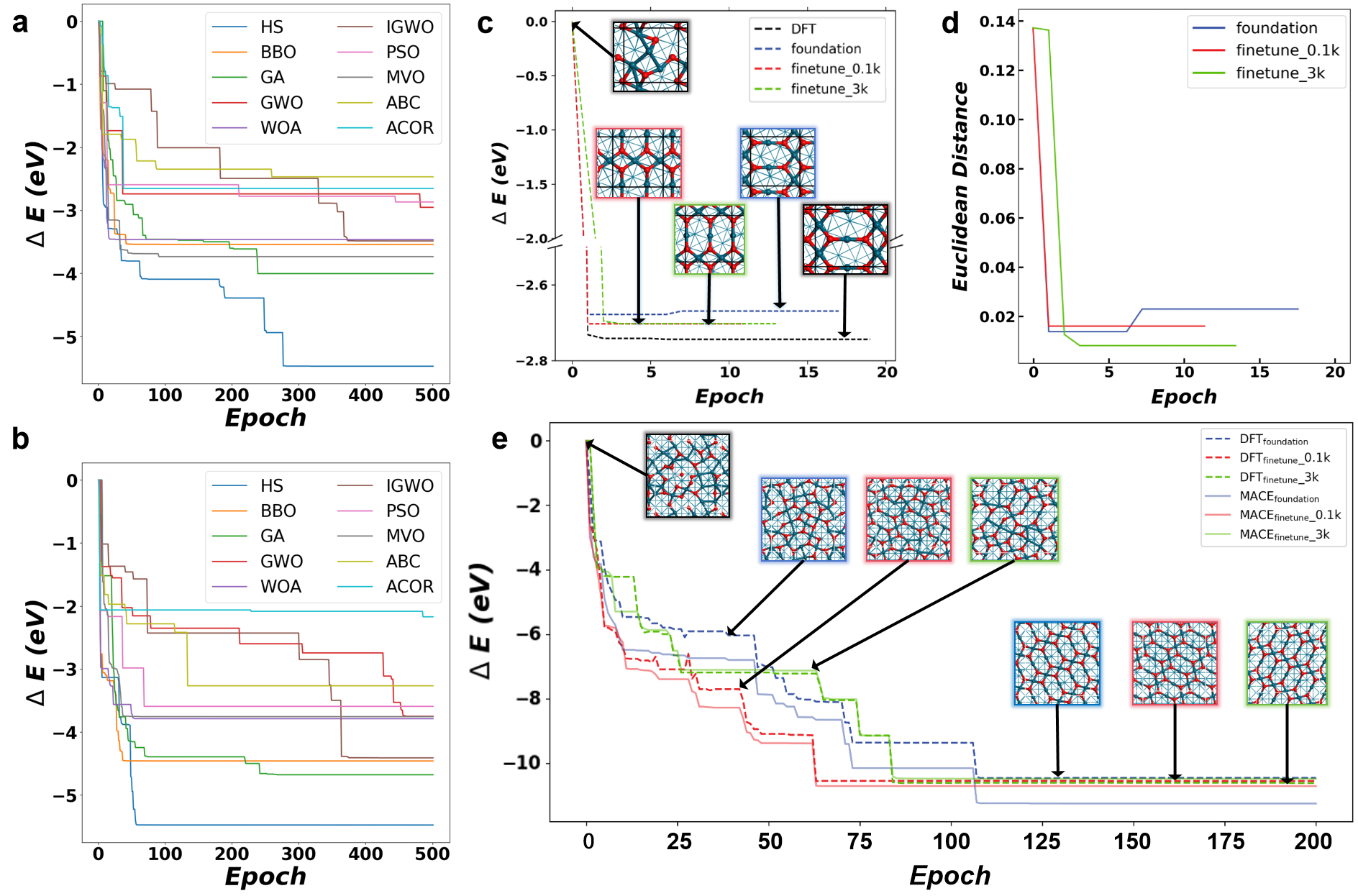}
  \caption{\textbf{HASGO algorithm benchmark and validation of the fine-tuning strategy.} Performance comparison of ten global optimisation algorithms on a Pd(100)-$p(4\times4)$ surface model over 500 epochs using the MACE-OMAT foundation model, with population sizes of 20 (\textbf{a}) and 50 (\textbf{b}) candidates per epoch. \textbf{c--e}, Robustness and scalability of the multi-head replay fine-tuning strategy for HASGO-driven Pd(100)-O surface reconstructions. \textbf{c} Energy evolution of HS-driven GO trajectories for a $p(\sqrt{5} \times \sqrt{5})$ surface, comparing DFT with three MLIP variants: u-MLIP, FT-0.1k, and FT-3k. \textbf{d.} Structural convergence quantified by the Euclidean distance of MACE-OMAT invariant descriptors relative to the DFT trajectory. \textbf{e}, Energy convergence for the extended $p(5 \times 5)$ system. In \textbf{c.} and \textbf{e.} dashed and solid lines denote DFT and MLIP energies, respectively, evaluated on the structures along each MLIP GO trajectory.}
  \label{fig:finetune-energy-comparison}
\end{figure}

A critical challenge in utilising MLIPs for GO is limited extrapolation capability in high-energy, OOD regions. To overcome this, we first evaluated our multi-head replay fine-tuning protocol using a smaller system of the Pd(100)-O $c(\sqrt{5} \times \sqrt{5})$ reconstruction, for which the entire GO process could be validated by DFT calculations. We compared the MACE-OMAT foundation model and two fine-tuned models trained on 0.1k and 3k samples against the standard HS+PBE-D3\cite{DFTD3_1,DFTD3_2} benchmark, respectively. Each search started from the same set of randomly generated, unrelaxed configurations to maintain consistency, with the HASGO algorithm executed over 20 epochs using a population size of 10 candidate structures per epoch. As illustrated in Fig.~\ref{fig:finetune-energy-comparison}c, all three MLIP variants exhibit rapid convergence behaviours that closely mirror the trajectory of the DFT for this small system. Notably, the 0.1k and 3k fine-tuned models converged to within $\sim$0.04 eV of the DFT benchmark, representing a massive throughput acceleration of $10^3$ relative to conventional first-principles methods. While the 0.1k model successfully identified the correct GM, the 3k model achieved higher geometric fidelity (Fig.~\ref{fig:finetune-energy-comparison}d), as quantified by Euclidean distances with the DFT benchmark in the MACE-OMAT descriptor space. This indicates that larger datasets provide the refined local force information necessary to drive the system into the exact DFT GM and yield more reliable dynamically oxidised surface structures.

When scaled to a complex Pd(100)-$p(5 \times 5)$ supercell, the exponentially expanded configurational space renders the full HASGO search at the DFT level computationally prohibitive. Thus, a series of DFT single-point calculations were performed based on the MLIP trajectories to benchmark the predictive confidence of MLIPs. Fig.~\ref{fig:finetune-energy-comparison}e shows that the foundation model exhibited a significant energy underestimation of $-0.80$ eV for the final GM structure compared with its DFT benchmark. The inclusion of just 100 fine-tuning samples ($\textrm{MACE}_\textrm{FT-0.1k}$) markedly reduced this deviation to $-0.16$ eV, which could compete with the inclusion of 3,000 fine-tuning samples ($\textrm{MACE}_\textrm{FT-3k}$, $+0.13$ eV). This indicates that our replay strategy is reliable for the identification of the GM even with sparse training data. However, notable energy deviations exist in the whole GO process using $\textrm{MACE}_\textrm{FT-0.1k}$, while $\textrm{MACE}_\textrm{FT-3k}$ exhibited almost consistent energies with its DFT benchmark. This demonstrates the ability of our replay strategy to construct a high-confidence PES using 3,000 fine-tuning samples. Interestingly, despite large energy deviations between the foundation model and two fine-tuned models based on the MLIP energies, the DFT results for the GM structures obtained by the three models are very close, indicating that HASGO is markedly robust for the search of the GM structure, even when the employed global PES exhibits energy deviations in certain regions. In summary, these results highlight a favourable trade-off between data efficiency and search performance, where a minimal sample set is sufficient to guide HASGO to the correct GM in previously inaccessible search spaces, while more samples are required to obtain a more reliable trajectory of the whole GO process.

\subsection*{Complex Surface Reconstructions}
Achieving an atomistic understanding of surface reconstructions under operating conditions still remains challenging. Within a GC ensemble at a fixed chemical potential of the reaction condition, identifying the globally optimized structure requires navigating an immense configurational space characterized by variable stoichiometry and symmetries. While surface reconstruction, particularly upon oxidation, has been extensively studied using small-scale, high-symmetry models, the limited knowledge of large-scale, low-symmetry phases due to structural complexity has hindered deeper atomic-level insights \cite{Limitsize1,Limitsize2,Limitsize3}. Having established the high-confidence PES provided by $\textrm{MACE}_\textrm{FT-3k}$, we deployed this model as the core computational engine for rapid energy evaluation. By coupling HASGO with the Automated Search for Optimal surface Phases (ASOPs) method\cite{ASOPs}, we sampled a vast configuration space spanning diverse unit cell dimensions and stoichiometric compositions to map the thermodynamically stable phases of oxidized metal surfaces under operating conditions. To validate the reliability of the HASGO framework, we first examined the structural evolution of oxidised Pd, Cu, and Ag surfaces under high-vacuum conditions relevant to atomic-scale microscopic characterisation (Supplementary Tables~\ref{tab:asop_pd}--\ref{tab:asop_ag_reaction}), benchmarking the predicted phases against experimental scanning tunnelling microscopy (STM) and high-resolution transmission electron microscopy (HRTEM) observations. We then extended HASGO to ethylene epoxidation on Ag(100) under realistic reaction condition, thereby assessing its capability for \textit{operando} structural simulation in an industrially relevant catalytic environment.

\subsubsection*{Phase evolution of PdO/Pd(100)}
\begin{figure}[H]
  \centering
  \includegraphics[scale=0.35]{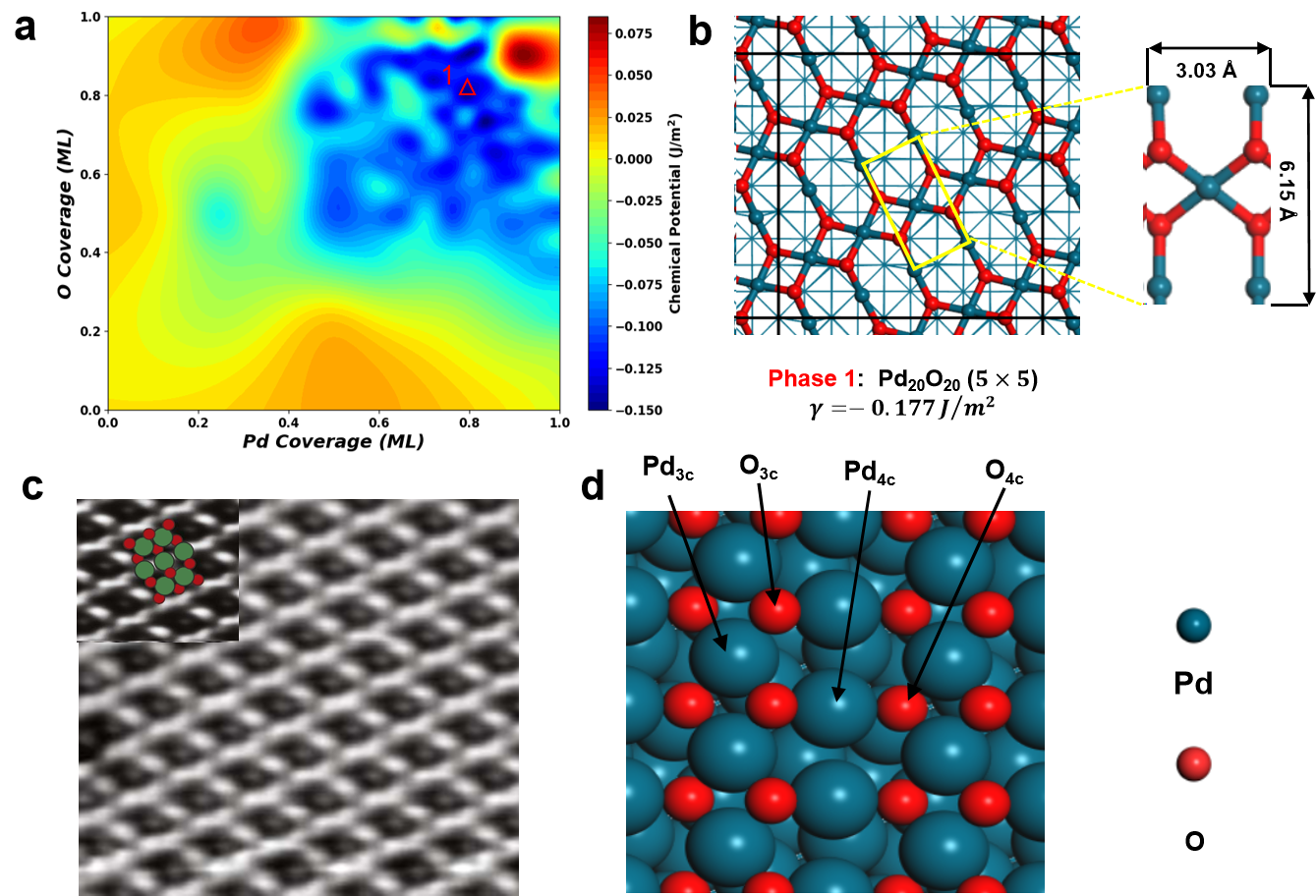}
  \caption{\textbf{The metal oxide surface reconstruction simulation and structural optimisation on Pd(100) under the STM imaging condition at 573~K and $P_\mathrm{O_2} = 5 \times 10^{-5}$~mbar\cite{PdOSTM}.} \textbf{a.} The PES contour map for palladium surface oxides on Pd(100) within the GC ensemble. \textbf{b,d} Top views of globally optimised structure in stick and CPK styles, respectively, identified as the most stable phase within the GC ensemble, featuring 0.8 ML Pd and 0.8 ML O on Pd(100)-$p(5 \times 5)$ through HASGO, as labelled in the PES map (Phase 1). \textbf{c.} Top-view STM image of $(\sqrt{5}\times\sqrt{5})$ R27$^\circ$ PdO(101) on Pd(100)\cite{PdOSTM}. Colour scheme: teal for Pd and red for O atoms.}
  \label{fig:Pd100}
\end{figure}

To map the structural landscape of Pd(100) under STM characterisation condition, we employed the HS algorithm to explore 44 Niggli-reduced unit cell types, 664 distinct chemical compositions, and the corresponding 492,800 local minima (Supplementary Table~\ref{tab:asop_pd} and Supplementary Fig.~\ref{sfig:PdO-asop}). The thermodynamic stability of each composition was quantified using the surface grand potential ($\gamma$, Eq.~\ref{Surface_Miu}) at $T = 573$~K and $P_{\mathrm{O_2}} = 5.0 \times 10^{-5}$~mbar, directly matching the experimental STM imaging condition\cite{PdOSTM}. Projection of these states onto the PES contour map identified the $(\sqrt{5} \times \sqrt{5})$ R27$^\circ$ PdO(101) surface overlayer on a $p(5 \times 5)$ supercell as the most thermodynamically stable phase (Fig.~\ref{fig:Pd100}a), with $\gamma = -0.177~\mathrm{J \cdot \text{m}^{-2}}$. By contrast, a control HASGO search constrained to a high-symmetry $p(2\times2)$ unit cell, which was frequently investigated\cite{p(2x2)-high-symmetry_1,p(2x2)-high-symmetry_2,p(2x2)-high-symmetry_3,p(2x2)-high-symmetry_4}, under the same thermodynamic condition recovered only a simple $(2\times2)$ phase with $\gamma = -0.082~\mathrm{J \cdot \text{m}^{-2}}$ (Supplementary Fig.~\ref{Sfig:Pd100-high-symmetry}), lying $0.095~\mathrm{J \cdot \text{m}^{-2}}$ above the variable-lattice optimum. This energetic separation shows that the PdO(101) overlayer is stabilised only when both the unit cell geometry and chemical composition are allowed to vary beyond the conventional $p(2\times2)$ constraint. The optimal phase corresponds to a $\text{Pd}_{20}\text{O}_{20}$ stoichiometry with symmetric 0.8 monolayer (ML) coverages of Pd and O (Fig.~\ref{fig:Pd100}b). Structurally, the predicted overlayer resolves into alternating parallel chains composed of distinct $[\text{PdO}_4]$ square-planar and $[\text{PdO}_2]$ linear coordination motifs (Fig.~\ref{fig:Pd100}d), in remarkable agreement with experimental STM topographies\cite{PdOSTM, PdO2002oxidation} (Fig.~\ref{fig:Pd100}c). Furthermore, the calculated lattice vectors of the predicted $\text{PdO}(101)$ overlayer give average short and long side lengths of $3.03\text{ \AA}$ and $6.15\text{ \AA}$, respectively, closely matching the corresponding values extracted from STM experiments\cite{PdO(101)-stm-vector-length}. Together, these results demonstrate that HASGO can resolve intricate, large-scale surface reconstructions without imposing prior structural bias.

\subsubsection*{Missing-row reconstruction of CuO/Cu(100)}
\begin{figure}[H]
  \centering
  \includegraphics[scale=0.35]{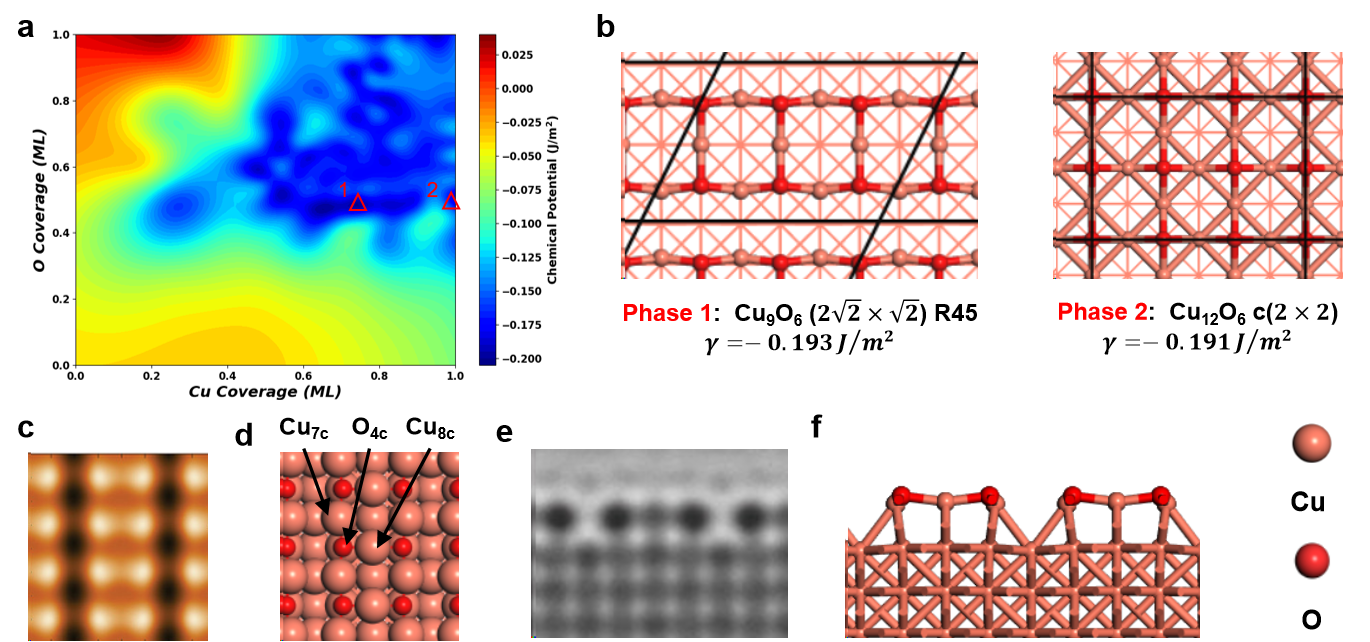}
  \caption{\textbf{The metal oxide surface reconstruction simulation and structural optimisation on Cu(100) under the STM imaging condition at $T = 673$~K and $P_{\mathrm{O_2}} = 7 \times 10^{-5}$~mbar.} \textbf{a.} The PES contour map for copper surface oxides on Cu(100) within the GC ensemble. \textbf{b.} Top view snapshots of two representative phases, as labelled in the PES map. Top views of \textbf{c.} experimental STM image for the $(2\sqrt{2}\times\sqrt{2})$ R45$^\circ$ CuO MRR monolayer on Cu(100)\cite{Cu100_MMR_top} and \textbf{d.} the most thermodynamically stable Phase 1. Side views of \textbf{e.} HRTEM image of reconstructed Cu(100) layers \cite{Cuside} and \textbf{f.} $(2\sqrt{2}\times\sqrt{2})$ R45$^\circ$ CuO MRR monolayer on Cu(100) of Phase 1. Colour scheme: coral for Cu and red for O atoms.}
  \label{fig:Cu100}
\end{figure}

We next applied the same search protocol to Cu(100) under STM imaging condition at $T = 673$~K and $P_{\mathrm{O_2}} = 7 \times 10^{-5}$~mbar\cite{Cuside}, exploring massive Niggli-reduced unit cells, chemical compositions, and local minima with the same number as those in the PdO/Pd(100) case (Supplementary Table~\ref{tab:asop_cu}). As illustrated in the phase stability diagram (Fig.~\ref{fig:Cu100}a), variable-lattice HASGO identified the $(2\sqrt{2} \times \sqrt{2})\text{R}45^\circ$ CuO missing-row reconstruction (MRR, Phase 1) as the thermodynamically stable overlayer, with $\gamma = -0.193~\mathrm{J \cdot \text{m}^{-2}}$. In contrast, the corresponding high-symmetry $p(2\times2)$ HASGO search yielded only a simple $(2\times2)$ phase with $\gamma = -0.176~\mathrm{J \cdot \text{m}^{-2}}$ (Supplementary Fig.~\ref{Sfig:Cu100-high-symmetry}), rendering the MRR inaccessible within the restricted cell description. This comparison confirms that the CuO missing-row phase emerges only when the search space is released from the symmetry and cell-size constraints imposed by the conventional $p(2\times2)$ model. This phase possesses a $\text{Cu}_{9}\text{O}_{6}$ stoichiometry (Fig.~\ref{fig:Cu100}b), and its structural parameters exhibit high fidelity to experimental HRTEM and STM benchmarks\cite{Cuside}. The structural integrity of this HASGO-optimised model is apparent in the side-view geometry (Fig.~\ref{fig:Cu100}f), which captures the characteristic buckling and elevated Cu--O rows intrinsic to MRR surfaces (Fig.~\ref{fig:Cu100}e)\cite{Cuside}. Plan-view analysis (Fig.~\ref{fig:Cu100}d) reveals that this reconstruction is driven by the periodic ejection of every fourth $[100]$ copper row. The resulting overlayer contains distinct coordination environments, in which surface O atoms occupy four-fold hollow-like sites ($\text{O}_{4\text{c}}$) and coordinate with undercoordinated Cu atoms ($\text{Cu}_{7\text{c}}$ and $\text{Cu}_{8\text{c}}$). This specific arrangement of Cu--O chains is expected to minimise surface strain and regulate the electronic properties of the oxidised Cu(100) interface\cite{CuO_MRR_theoritical}. Capturing this low-symmetry MRR phase therefore highlights the capacity of HASGO to overcome the rigid symmetry constraints of small unit cells, which can otherwise bias traditional modelling against real active phases.
\par
Phase 2 is the second most thermodynamically stable structure in the same phase diagram (Fig.~\ref{fig:Cu100}a), featuring a $\text{Cu}_{12}\text{O}_{6}$ stoichiometry with Cu and O coverages of 1.0 and 0.5~ML, respectively (Fig.~\ref{fig:Cu100}b). With $\gamma = -0.191~\mathrm{J \cdot \text{m}^{-2}}$, Phase 2 lies only $0.002~\mathrm{J \cdot \text{m}^{-2}}$ above the GM of Phase 1. This marginal energetic penalty is consistent with its established role as the experimental and theoretical structural precursor to the full $(2\sqrt{2}\times\sqrt{2})\text{R}45^\circ$ missing-row reconstruction\cite{Cu100_c2x2->MMR_experiment_1,Cu100_c2x2->MMR_experiment_2,Cu100_c2x2->MMR_theoritical_1,Cu100_c2x2->MMR_theoritical_2}.

\subsubsection*{Missing-row reconstruction of AgO/Ag(100)}
Variable-lattice HASGO was further extended to Ag(100) (Supplementary Table~\ref{tab:asop_ag}) under the reported in situ X-ray diffraction (XRD) characterisation condition of $T = 440$~K and $P_{\mathrm{O_2}} = 1 \times 10^{-1}$~mbar\cite{AgO-MRR-1}. As displayed in the PES contour map (Fig.~\ref{fig:Ag100}a), HASGO identified the AgO missing-row reconstruction (MRR, Phase~1) as the thermodynamically stable surface phase (Fig.~\ref{fig:Ag100}b), with $\gamma = -0.125~\mathrm{J \cdot \text{m}^{-2}}$. Under the same thermodynamic condition, the restricted $p(2\times2)$ HASGO search recovered a simple $(2\times2)$ phase with $\gamma = -0.090~\mathrm{J \cdot \text{m}^{-2}}$ (Supplementary Fig.~\ref{Sfig:Ag100-high-symmetry}), $0.035~\mathrm{J \cdot \text{m}^{-2}}$ higher than the variable-lattice optimum. The energetic preference for the larger, lower-symmetry phase reinforces that the AgO MRR motif is a genuine outcome of symmetry-unconstrained GO. Notably, this MRR motif on Ag(100) is structurally analogous to the CuO MRR identified on Cu(100), featuring periodic removal of metal rows that generates alternating ridges and troughs on the surface. Detailed inspection of the surface architecture (Fig.~\ref{fig:Ag100}c) shows that the reconstruction is characterised by Ag--O--Ag chains along the remaining rows, with surface O atoms occupying four-fold hollow-like sites and coordinating to undercoordinated Ag atoms at the ridge positions. These structural features have been identified experimentally by both in situ XRD\cite{AgO-MRR-1} and high-resolution electron-energy-loss spectroscopy\cite{AgO-MRR-2}. This arrangement of Ag--O chains effectively relieves surface tensile stress and stabilises the oxidised Ag(100) interface, analogous to the mechanism observed for Cu(100).

\subsubsection*{O$_5$ reconstruction of AgO/Ag(100) during ethylene epoxidation}
Inspired by the excellent agreement between HASGO predictions and experimental observations under high-vacuum conditions, we further applied variable-lattice HASGO to the ethylene epoxidation over Ag(100) under the realistic reaction condition of $T = 500$~K, $P_{\mathrm{O_2}} = 1$~bar, and $P_{\mathrm{C_2H_4}} = 1$~bar.  Under these operating conditions, the thermodynamically relevant surface phase space is no longer governed solely by Ag--O stoichiometry, as ethylene--surface interactions can affect the surface grand potential as well, thereby modifying the surface atomic configuration and reshaping the surface free energy landscape and relative stability of reconstructed Ag--O overlayers\cite{Surfacereconstructions,lzp_Ag4OAg,Adsorbate-induced_reconstruction_1,Adsorbate-induced_reconstruction_2}. We therefore treated ethylene adsorption and Ag--O reconstruction within the same variable-lattice search, rather than optimising the oxide surface first and adding ethylene only afterwards.

\begin{figure}[H]
  \centering
  \includegraphics[scale=0.35]{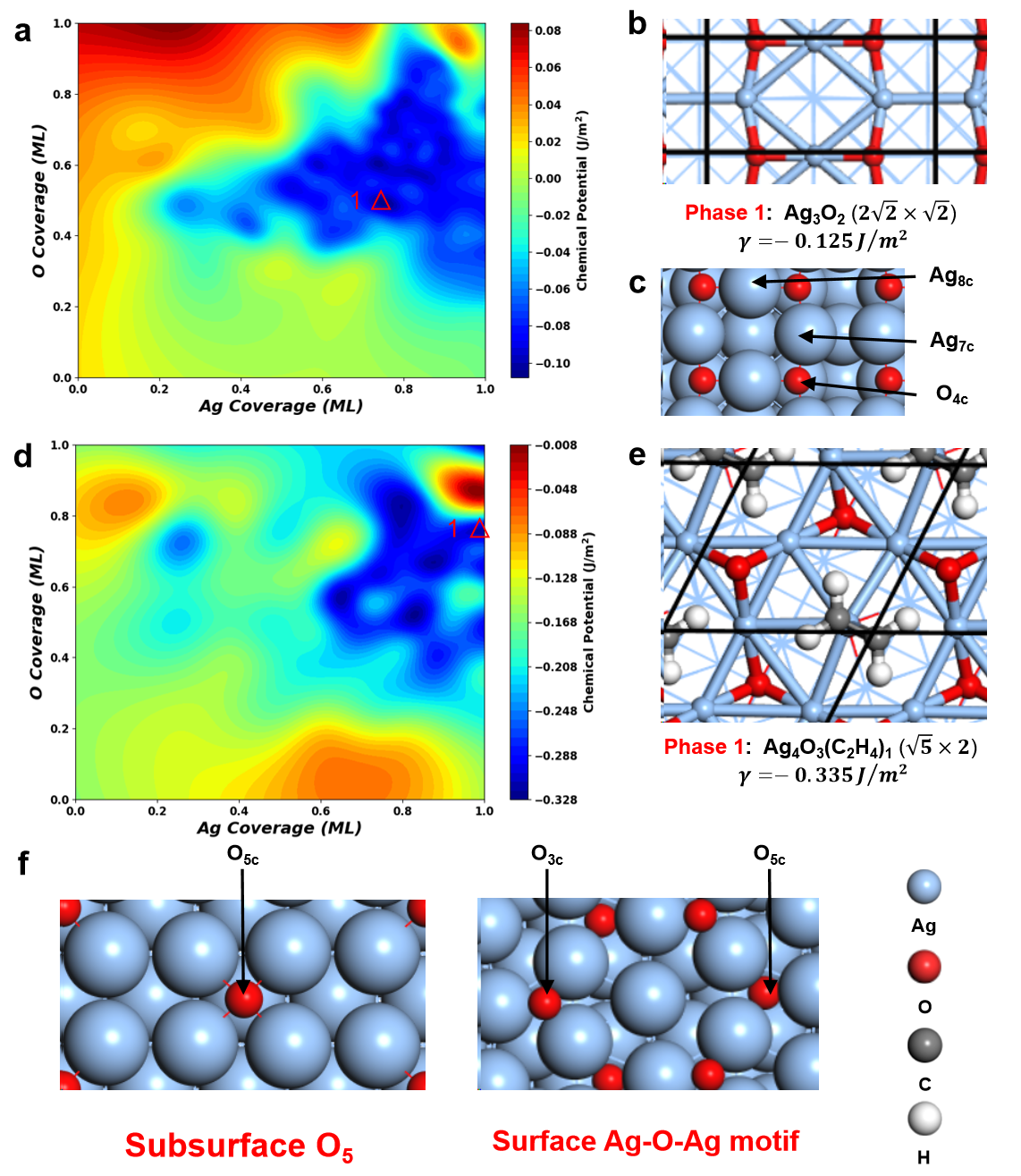}
  \caption{\textbf{The Ag surface structure evolution under various conditions.} \textbf{a.} The PES contour map for silver surface oxides on Ag(100) at 440~K and $P_\mathrm{O_2} = 1 \times 10^{-1}$~mbar from ASOP simulation. \textbf{b,c} Top view snapshot of globally optimised phase, identified as the most stable phase within the GC ensemble, featuring 0.75~ML Ag, 0.5~ML O on Ag(100), as labelled in the PES map (a, Phase 1). \textbf{d.} The PES contour map for silver surface oxides on Ag(100) at 500~K and 1~bar of both oxygen and ethylene from ASOP simulation. \textbf{e.} Top view globally optimised structure, identified as the most stable phase within the GC ensemble, featuring 1.0~ML Ag, 0.75~ML O and 0.25~ML ethylene on Ag(100), as labelled in the PES map (d, Phase 1). \textbf{f.} Top view of the subsurface and surface AgO layer in Phase 1. Colour scheme: blue for Ag, red for O, grey for C and white for H atoms.}
  \label{fig:Ag100}
\end{figure}

The relative thermodynamic stability of the ternary Ag--O--C$_2$H$_4$ phases was evaluated using a $\gamma$ analogous to that used for the PdO, CuO, and AgO systems (Eq.~\ref{Surface_Miu}). For the reaction system, this formulation additionally includes the chemical potential of ethylene ($\mu_{\mathrm{C_2H_4}}$) and the adsorption free energy of ethylene ($G_{\mathrm{C_2H_4}, {\mathrm{ads}}}$). Ethylene coverage was therefore explicitly included as a compositional variable, enabling Ag--O structures with different C$_2$H$_4$ loadings to be compared within a single GC ensemble (Supplementary Table~\ref{tab:asop_ag_reaction}). This treatment allows the phase search to capture adsorbate-driven changes in the stability of reconstructed Ag--O overlayers under ethylene epoxidation condition. As displayed in the PES contour map (Fig.~\ref{fig:Ag100}d), variable-lattice HASGO identified the AgO overlayer with O$_5$ (Phase 1) as the GM during ethylene epoxidation, featuring 1.0~ML Ag, 0.75~ML O, and 0.25~ML ethylene ($\textrm{Ag}_4\textrm{O}_3(\textrm{C}_2\textrm{H}_4)_1$, Fig.~\ref{fig:Ag100}e), with $\gamma = -0.335~\mathrm{J \cdot \text{m}^{-2}}$. This $\gamma$ is substantially lower than that of the ethylene-free variable-lattice AgO/Ag(100) optimum at the same oxygen chemical potential ($\gamma = -0.161~\mathrm{J \cdot \text{m}^{-2}}$, Supplementary Fig.~\ref{Sfig:AgO-operating}), corresponding to a stabilisation of $0.174~\mathrm{J \cdot \text{m}^{-2}}$ upon inclusion of ethylene. This energetic separation indicates that ethylene adsorption does not merely decorate a pre-existing Ag--O overlayer, but actively reshapes surface phase stability and promotes a more favourable adsorbate-coupled reconstruction. As shown in Fig.~\ref{fig:Ag100}f, the predicted structure is decorated with chemisorbed ethylene on Ag atoms located on a nearly hexagonal Ag--O--Ag skeleton and above the subsurface O$_5$ species, namely [Ag$_4$OAg]. This geometry indicates that ethylene adsorption and oxide reconstruction are coupled in the thermodynamically preferred \textit{operando} phase. The reconstructed Ag--O skeleton provides exposed Ag sites for ethylene adsorption, while the same overlayer accommodates the square-pyramidal subsurface O$_5$ environment as its characteristic oxygen motif. Importantly, this assignment is consistent with recent infrared spectroscopic discovery\cite{lzp_Ag4OAg} although it could not be directly observed by STM or HRTEM due to pressure gap for the operating condition. This identified O$_5$ motif as the most thermodynamically stable structure under ethylene epoxidation condition. When the same reaction environment is simulated within a high-symmetry $p(2\times2)$ unit cell, the search converges to a ($2\times2$) phase with $\gamma = -0.250~\mathrm{J \cdot \text{m}^{-2}}$ (Supplementary Fig.~\ref{Sfig:Ag100-ethene}), which remains $0.085~\mathrm{J \cdot \text{m}^{-2}}$ less stable than the variable-lattice O$_5$ phase and entirely overlooks the adsorbate-coupled subsurface O$_5$ motif. This comparison underscores that the low-symmetry, large-scale exploration enabled by HASGO is indispensable for identifying the authentic \textit{operando} active phase under realistic catalytic environments.

\section*{Discussion}
In this work, we introduced HASGO, a highly efficient framework designed to resolve the atomic-level structural evolution of complex surface reconstructions under operating conditions. By integrating the HS algorithm with a multi-head replay fine-tuning strategy for u-MLIPs, HASGO mitigates PES softening, a pervasive limitation in u-MLIPs when evaluating OOD atomic configurations far from equilibrium. Notably, fine-tuning with a sparse dataset of merely 100 configurations is sufficient to guide the algorithmic search to the correct GM, while utilising 3,000 samples achieves near-exact energetic and topological agreement with DFT benchmarks. Concurrently, the inherent fault-tolerant mechanism and the balance between configurational exploration and local exploitation adapt the HS algorithm for use with MLIPs, exhibiting superior performance in finding the most thermodynamically stable structures with competitive computational efficiency. These advances extend the role of MLIP-accelerated GO beyond computational speed-up, enabling systematic exploration of variable unit cells and compositions regions, covering a great number of large, low-symmetry slabs, within the GC phase space that are essential for resolving realistic catalyst surfaces under operating conditions.

The surface-reconstruction results demonstrate the chemical consequence of this expanded search capability. By allowing stoichiometry, cell geometry, and long-range ordering to vary within the GC ensemble, variable-lattice HASGO recovers the experimentally observed $p(5\times5)$ PdO(101) overlayer on Pd(100), the $(2\sqrt{2}\times\sqrt{2})$ R45$^{\circ}$ CuO missing-row reconstruction on Cu(100), and the analogous AgO MRR phase on Ag(100), whereas searches on conventional high-symmetry $p(2\times2)$ unit cell converge to simpler structures with higher surface grand potentials. The agreement with STM, HRTEM, and XRD observations therefore validates more than the predicted atomic geometries. It shows that experimentally observed oxide overlayers can be understood as thermodynamic outcomes of a broader GC landscape that must be sampled without imposing restrictive symmetry assumptions. This requirement becomes even more pronounced under ethylene epoxidation condition. When ethylene coverage is incorporated into the phase composition, variable-lattice HASGO identifies an adsorbate-coupled Ag--O reconstruction containing the square-pyramidal subsurface O$_5$ motif, consistent with the operando phase recently inferred from infrared spectroscopy. The progression from high-vacuum Pd, Cu, and Ag oxide reconstructions to the Ag--O--C$_2$H$_4$ reaction environment demonstrates that HASGO does not simply reproduce known structural motifs, but enables the thermodynamic evolution of the active phase to be followed as the chemical environment approaches catalytic operation. By combining data-efficient u-MLIP fine-tuning, fault-tolerant HS exploration, and first-principles surface thermodynamics, HASGO renders low-symmetry, compositionally flexible active phases computationally accessible, providing a practical basis for mechanistic understanding and rational design of dynamic catalytic interfaces.

\section*{Methods}
\subsection*{u-MLIP Model Fine-tuning}
\label{subsec:finetune}
The fine-tuning component of HASGO employs the multi-head replay training strategy. To execute this strategy, a system-specific dataset pool must first be constructed, utilising MACE-OMAT\cite{omat24} as the foundation model. Exhaustive GM searches are conducted on small-scale targets using HASGO. Importantly, the complete local optimisation trajectories generated during these searches are retained. This ensures the incorporation of diverse configurations, including high-energy, high-force, and OOD states, which are vital for mitigating systematic PES softening\cite{EMLP}. Unphysical structures, such as those featuring uncharacteristically short interatomic distances, are automatically filtered out prior to sampling.

To compile the final fine-tuning dataset, configurations are iteratively sampled from the pool using the FPS method within the high-dimensional MACE descriptor space, accelerated by a KDTree-based greedy search\cite{kd-tree}. This maximin distance approach maximises geometric dispersion, guaranteeing comprehensive coverage of the relevant configurational space while minimising the computational burden of DFT labelling. To evaluate the effect of dataset volume, variants containing 0.1k and 3k samples were benchmarked (detailed in Supplementary Section~\ref{sec:multi-head-replay-finetune-setting}). To reconcile the trade-off between data efficiency of OOD samples and the utilisation of the broad pre-trained knowledge embedded in the u-MLIP, we employed a multi-head replay fine-tuning strategy. A complementary replay dataset comprising 10,000 representative structures was sampled from the broader OMAT dataset using FPS. Combining both the specified and replay datasets during training yields a robust, system-tailored u-MLIP without catastrophic forgetting (detailed in Supplementary Section~\ref{sec:multi-head-replay-finetune-setting}).

\subsection*{Harmony Search Method}
The HS algorithm is a metaheuristic optimisation technique inspired by musical improvisation\cite{HS}. As illustrated in Fig.~\ref{Sfig:HS}a, for atomic structural GO, the process initialises by populating a harmony memory ($\mathbf{HM}$). The $\mathbf{HM}$ functions as a repository of candidate solution vectors (harmonies, $\mathbf{X}$) and their corresponding post-relaxation fitness values (energies, $\mathbf{E}$). Its fundamental structure is as follows:

\begin{equation}\label{HM}
\mathbf{HM} = \left[\begin{array}{c|c}
\mathbf{X}_1 &  E_1 \\
\mathbf{X}_2 &  E_2 \\
\vdots &  \vdots \\
\mathbf{X}_{M} & E_{M}
\end{array}\right] =
\left[\begin{array}{ccccccc|c}
{x_{11}} & {y_{11}} & {z_{11}} & \ldots & {x_{1N}} &   {y_{1N}} & {z_{1N}} & E_1 \\
{x_{21}} & {y_{21}} & {z_{21}} & \ldots & {x_{2N}} &   {y_{2N}} & {z_{2N}} & E_2 \\
\vdots & \vdots & \vdots & \ddots & \vdots & \vdots & \vdots \\
{x_{M1}} & {y_{M1}} & {z_{M1}} & \ldots & {x_{MN}} &   {y_{MN}} & z_{MN} & E_{M}
\end{array}\right]
\end{equation}

Each harmony, $\mathbf{X}_{i}=(x_{i1}, y_{i1},z_{i1},...,x_{iN},y_{iN},z_{iN})$, is a $3N$-dimensional vector representing the flattened Cartesian coordinates of the $N$ atoms. To model surface reconstructions accurately, the periodic slab is partitioned along the $z$-axis into vacuum, surface, buffer, and substrate regions (Supplementary Fig.~\ref{Sfig:HS}b). During optimisation, HS actively updates the surface coordinates, buffer atoms undergo local relaxation only via the LBFGS algorithm, and substrate atoms remain rigidly fixed.

The optimisation trajectory, termed ``improvisation,'' iteratively updates the configurations stored within the HM. To guarantee expansive PES exploration, initial configurations ($\mathbf{HM}(0)$) are generated with randomly displaced surface atoms in the surface region and locally relaxed (see Supplementary Section~\ref{sec:initial_state}). The improvisation process is governed by three primary parameters---harmony memory consideration rate (HMCR), pitch adjustment rate (PAR), and bandwidth (BW)---to dynamically balance between random exploration for a novel candidate and exploitation of the candidate with the best fitness stored in the previous $\mathbf{HM}$.

During memory consideration, HMCR dictates whether the algorithm exploits known states or explores new ones:

\begin{equation}\label{HMCR}
\mathbf{X}_i(n) (i \in 1,2, \cdots, M) = 
\begin{cases} 
\mathbf{X}^{\mathrm{best}}(n-1) & \mathrm{if\ } r_{i} < \mathrm{HMCR} \\[6pt]
\mathbf{X}^{\mathrm{rand}}_i & \mathrm{if\ } r_{i} \geq \mathrm{HMCR}
\end{cases}
\end{equation}

Here, $\mathbf{X}_i(n)$ is the $i^{\mathrm{th}}$ candidate at epoch $n$, $\mathbf{X}^{\mathrm{best}}(n-1)$ the lowest-energy structure from the preceding epoch, and $\mathbf{X}^{\mathrm{rand}}_i$ denotes a novel, randomly generated and locally relaxed structure. A uniform random number $r_{i}$ is evaluated against HMCR to resolve this probabilistic choice.

Following a random choice between exploration and exploitation, candidates undergo structural perturbation (pitch adjustment and refinement), providing fault tolerance against marginal discrepancies between the MLIP and DFT PES. If a random number $r_{ik}$ is less than PAR, the positional vector $X_{ik}$ is perturbed:

\begin{equation}\label{PAR}
\mathbf{X}_{ik}(n) = 
\begin{cases}
\mathbf{X}_{ik}(n) \pm \mathrm{BW}(n) \cdot \mathcal N (1/2, 1/36) &  \mathrm{if\ } r_{ik}< \mathrm{PAR}, \\[6pt]
\mathbf{X}_{ik}(n) &  \mathrm{if\ } r_{ik} \geq \mathrm{PAR}
\end{cases}
\end{equation}

where $i \in \{1,2, \ldots, M\}$, $k \in \{x_{1}, y_{1}, z_{1}, \ldots, x_{N}, y_{N}, z_{N}\}$, and $\mathcal N (1/2, 1/36)$ is a Gaussian distribution mainly between $[0, 1]$. The bandwidth BW governs the maximum displacement distance of atomic coordinates, and PAR governs the likelihood of locally refining the atomic coordinate during improvisation. During the refinement stage, BW is decreased over epochs via an exponential decay scheme to narrow the search radius, facilitating a vital algorithmic transition from exploration-first to exploitation-first. This enables the HS to perform intensive, high-resolution searches within confined and high-fitness regions, ultimately enhancing the efficiency of locating the GM structure.

\subsection*{Metal Oxide Surface Reconstruction Simulation under Grand Canonical Ensemble.}
\label{sec:asop}
To rigorously assess surface oxide formation on Pd(100), Cu(100) and Ag(100) under high-vacuum oxidation conditions, and Ag(100) under ethylene epoxidation condition, we integrated HASGO with the ASOPs framework\cite{ASOPs} for simulation in a GC ensemble. This automated pipeline facilitates composition grid generation, structural exploration, and thermodynamic phase selection via top-$k$ algorithms. The searchable compositional domain included all Niggli reduced cells with a lattice vector ratio $(a/b)$ strictly less than 3, accommodating surface metal and oxygen coverages between 0 and 1 monolayer (ML), while surface C$_2$H$_4$ coverage was 0 for high-vacuum oxidation conditions and 0.01--0.33 ML for ethylene epoxidation condition. For the Ag--O--C$_2$H$_4$ reaction system, ethylene molecules and the variable Ag--O surface atoms were included in the same active region and globally optimised concurrently at each sampled composition, rather than treating ethylene as a post-local-optimisation adsorbate on a preformed oxide surface. To maintain tractability, phase exploration per grid was capped at 15 distinct chemical compositions.

We utilised a hierarchical search strategy to efficiently map this expansive configuration space (Supplementary Figs.~\ref{sfig:PdO-asop}--\ref{sfig:AgO-ethene-asop}). Broad screening was initially executed on smaller unit cells with surface area ratios ranging from 4 to 12. The HS algorithm, driven by the fine-tuned u-MLIP, evaluated the GM energy for each composition. The surface grand potential ($\gamma$) of these small-cell structures was then calculated and interpolated to generate a preliminary thermodynamic contour map. Directed by this initial PES mapping, we expanded our structural search to larger supercells ($p(4\times4)$ and $p(5\times5)$) for the most promising compositions. This subsequent exploration transcends the symmetry constraints of smaller cells, allowing for the discovery of intricate surface reconstructions. The high-resolution data from these large-scale simulations subsequently refined the initial PES contours, providing an accurate thermodynamic description.

The relative thermodynamic stability of the identified phases was evaluated using the surface grand potential ($\gamma$), formulated as:

\begin{equation}\label{Surface_Miu}
\gamma = \frac{E_{\mathrm{M}_{x}\mathrm{O}_{y}(\mathrm{C_2H_4})_{z}} - E_{\mathrm{surf}} - x \mu_{\mathrm{M}} - y \mu_{\mathrm{O}} - z (\mu_{\mathrm{C_2H_4}} - G_{\mathrm{C_2H_4}, {\mathrm{ads}}})}{A}
\end{equation}

where M indicates the underlying transition metal; $E_{\mathrm{M}_{x}\mathrm{O}_{y}(\mathrm{C_2H_4})_{z}}$ and $E_{\mathrm{surf}}$ represent the calculated energy of the globally optimised phase and the pristine metallic surface, respectively; $\mu_{\mathrm{M}}$, $\mu_{\mathrm{O}}$, and $\mu_{\mathrm{C_2H_4}}$ represent the chemical potentials of the metal atom, O atom, and ethylene molecule, respectively (details in Supplementary Section~\ref{sec:Gibbs_free_energy}); $x$, $y$, and $z$ denote the numbers of metal, O atoms, and ethylene molecules within the surface oxide layer; $G_{\mathrm{C_2H_4}, {\mathrm{ads}}}$ represents the thermal effects of the adsorbed ethylene, including the adsorbate entropies and enthalpies; and $A$ corresponds to the surface area of the unit cell.

\section*{Data and Code Availability}

The code and datasets generated during and/or analysed during the current study will be made publicly available upon the formal publication of this manuscript.

\section*{Acknowledgements}
This work was supported by the National Key R\&D Project of China (2023YFA1507601). We acknowledged the Open Source Supercomputing Center of S-A-I for providing the computing resources

\subsection*{Corresponding authors}
Correspondence to Xiao-Ming Cao.

\section*{Author Contributions}

S.C. developed the HASGO framework, performed the calculations, and analysed the data. X.-M.C. conceived and supervised the project. Both authors discussed the results and contributed to writing the manuscript.

\section*{Competing Interests}

The authors declare no competing interests.

\section*{Additional Information}

\textbf{Supplementary Information} is available for this paper.

\textbf{Correspondence} and requests for materials should be addressed to X.-M.C. (\texttt{xmcao@sjtu.edu.cn}).

\newpage
% SI
{\centering\Titlefont\textit{Supplementary Information} for ``Scalable Prediction of Complex Surface Reconstructions under Operating Conditions via Harmony-Search-Based Global Optimization''\par}

\vspace{20pt}

% \begin{document}
\appendix
% \maketitle

% \vspace{-1em}
% \begin{center}
% $^{1}$State Key Laboratory of Green Chemical Engineering and Industrial Catalysis, Centre for Computational Chemistry and Research Institute of Industrial Catalysis, East China University of Science and Technology, Shanghai 200237, China.\\
% $^{2}$State Key Laboratory of Synergistic Chem-Bio Synthesis, School of Chemistry and Chemical Engineering, Shanghai Jiao Tong University, Shanghai 200240, China.\\[0.5em]
% $^{\ast}$Corresponding author: \texttt{xmcao@sjtu.edu.cn}
% \end{center}

% \tableofcontents
\renewcommand{\thesection}{S\arabic{section}}  
\renewcommand{\thetable}{S\arabic{table}}  
\renewcommand{\thefigure}{S\arabic{figure}}
\setcounter{figure}{0}
\setcounter{table}{0}
\setcounter{section}{0} 

\section{Initial Structure Generation}
\label{sec:initial_state}
HASGO is designed to operate autonomously, automatically generating the initial atomic structure when no user-defined configuration is provided. To automatically generate the initial surface structure, the p(1 $\times$ 1) primitive cell for the specified (hkl) surface, defined by a [$u$, $v$] vector, can be cleaved from the bulk crystal. Any supercell of the surface defined by [$u'$, $v'$] is first constructed through positive definite transition matrix transformation. The surface area $A$ is then explicitly calculated as the magnitude of the cross product of these two vectors, $A$ = $|u' \times v'|$. This geometric parameter represents the total area of the transformed unit cell and serves as the essential normalizing factor in our thermodynamic calculations. By deriving A directly from the supercell vectors, we ensure that the surface grand potential is accurately scaled, allowing for a consistent stability comparison across different cell sizes and symmetries. This transformation is fundamentally guided by the principles of the Niggli reduced cell theory$^{\text{\small\cite{niggli}}}$, as presented in Supplementary Fig.~\ref{Sfig:niggli}. 
\par
Subsequently, the atoms designated for global optimization are sequentially placed at random adsorption sites via Delaunay triangulation on the substrate within the predefined spatial boundaries, as presented in Supplementary Fig.~\ref{Sfig:initial_state_surface}. To ensure physical consistency during global optimization process, the system employs a 4-layer periodic slab model (as illustrated in Supplementary Fig.~\ref{Sfig:HS}b): the bottom atomic layer is fixed, the intermediate 2 buffer layers are restricted to local relaxation, while the active surface region is subjected to global and local optimization via the Harmony Search algorithm.

\begin{figure}[H]
  \centering
  \includegraphics[scale=0.75]{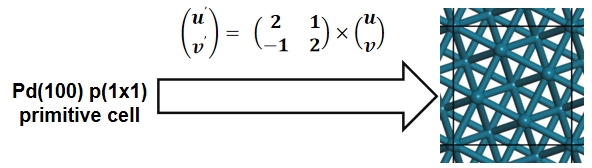}
  \caption{\textbf{Examples of the methodology of composition grid generation}. Generating the grid of ($\sqrt{5}\times\sqrt{5}$) reconstructed Pd(100) is illustrated.}
  \label{Sfig:niggli}
\end{figure}

\begin{figure}[H]
  \centering
  \includegraphics[scale=0.4]{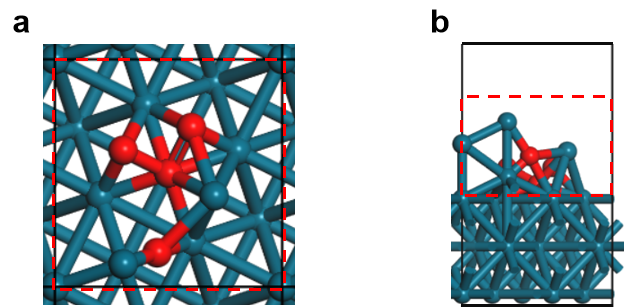}
  \caption{\textbf{Examples of automatically generated surface initial state.} \textbf{a.} Top and \textbf{b.} side view of ($\sqrt{5}\times\sqrt{5}$) reconstructed Pd(100) initial state is illustrated. Red dashed lines refers to the boundary for HASGO global optimization. Dark blue atoms for Pd, red atoms for O.}
  \label{Sfig:initial_state_surface}
\end{figure}

\newpage
\section{Multi-head Replay Finetuning Setting}
\label{sec:multi-head-replay-finetune-setting}
Settings and hyperparameters of the multi-head replay finetuning on MACE-OMAT-0-medium model foundation model. In this study, the MACE$^{\text{\small\cite{MACE}}}$ model was utilized to fit the MACE-OMAT foundation model. Two datasets were constructed for training: a target dataset containing system-specific structures with a 90:10 train-validation split, and a replay buffer of 10,000 structures sampled from the OMAT24$^{\text{\small\cite{omat24}}}$ dataset using the Farthest Point Sampling (FPS) method with pseudo-label replay enabled. The model architecture consisted of 2 interaction blocks with 128 channels per block, maximum angular momentum L$_{max}$ was set as 1, and correlation order of 3. A cutoff radius of 6.0~\AA\ was employed for interatomic interactions. Training was performed with a batch size of 16 for 200 epochs using exponential moving average (EMA) and stochastic weight averaging (SWA) techniques. Loss function weights were set to 1000, 100, and 10 for forces, energies, and stresses, respectively. The total training loss ($\mathcal{L}_{\text{total}}$) was defined as a weighted sum of the loss calculated on the finetuning dataset ($\mathcal{L}_{\text{ft}}$) and the pretraining dataset ($\mathcal{L}_{\text{pt}}$), setting weight 10.0, 1.0 for finetuning dataset and pretraining dataset respectively. The learning rate scheduler employed a patience of 15 epochs with early stopping at 30 epochs. All calculations were conducted with a random seed of 123 for reproducibility. All remaining hyperparameters and settings utilized during model training and evaluation were maintained at the default values established in MACE (https://github.com/ACEsuit/mace).
\par
To construct the "Specified Dataset" in Method \ref{subsec:finetune}, we designed two comparative sub-datasets containing 0.1k and 3k samples, respectively, to evaluate the model's sensitivity to data density. Initially, we use the Farthest Point Sampling (FPS) method to select 5k representative samples from the comprehensive pool of local optimization trajectories. To refine these samples for effective fine-tuning, we applied stringent physical constraints based on energy and force thresholds. For the 0.1k dataset, we excluded all structures with energy $>$ 0 eV or maximum atomic force $>$ 10.0~$\mathrm{eV \, \text{\AA}^{-1}}$, while further thinning the dataset by removing 90\% of configurations with maximum forces greater than 1.0~$\mathrm{eV \, \text{\AA}^{-1}}$. For the 3k dataset, we excluded all structures with energy $>$ 0 eV or maximum atomic force $>$ 100.0~$\mathrm{eV \, \text{\AA}^{-1}}$, while further thinning the dataset by removing 90\% of configurations with maximum forces greater than 10.0~$\mathrm{eV \, \text{\AA}^{-1}}$. All energy, force, and descriptor calculations were performed using the MACE-OMAT foundation model. 

\newpage
\section{Harmony Search Setting}
\label{sec:HS_setting}
\begin{figure}[H]
  \centering
  \includegraphics[scale=0.2]{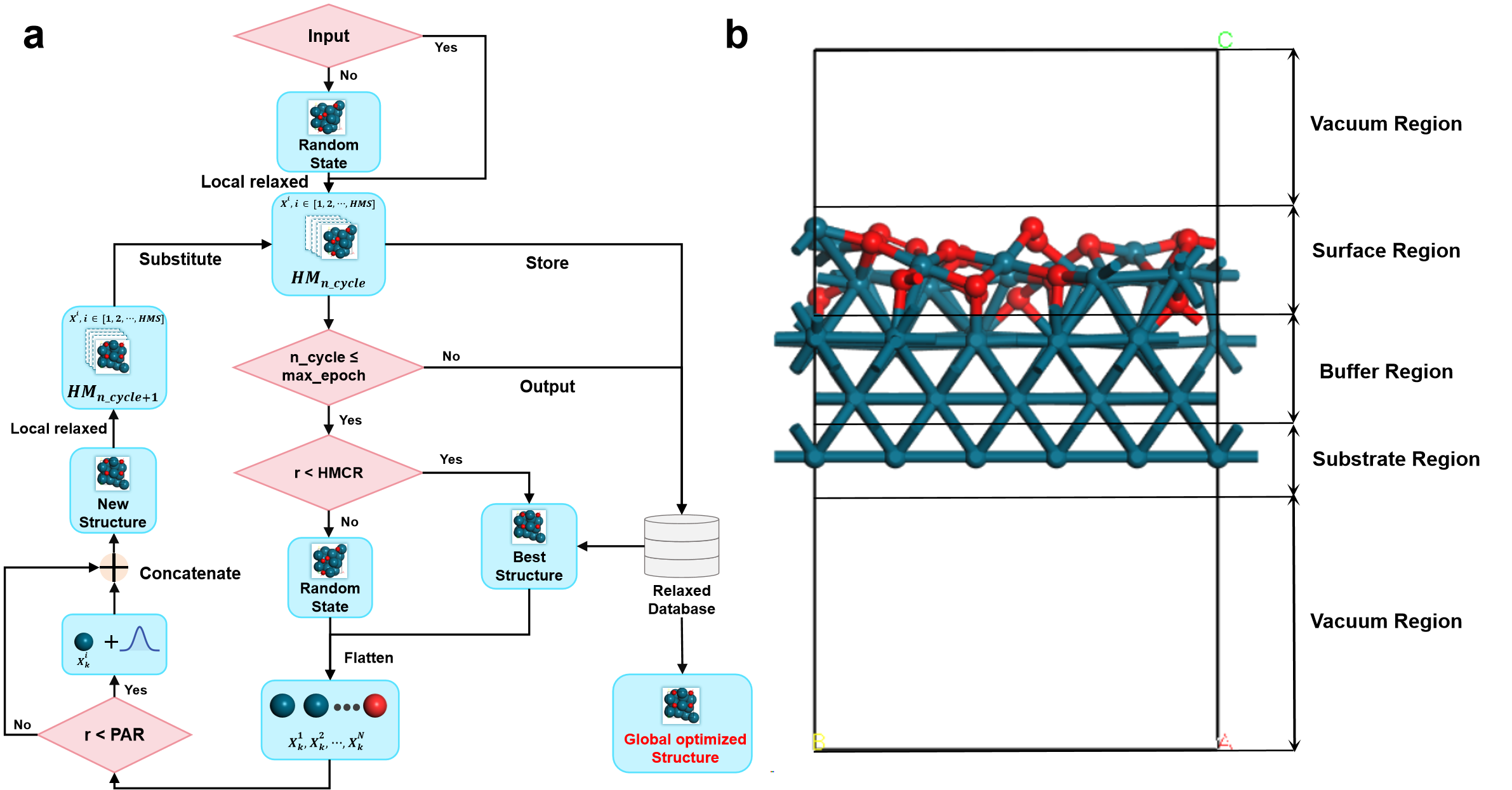}
  \caption{\textbf{The application of the Harmony Search (HS) algorithm in the global optimisation task.} \textbf{a.} Illustration of the operating principle of the HS algorithm in HASGO. \textbf{b.} Side view of the periodic slab model employed for global optimisation by HASGO.}
  \label{Sfig:HS}
\end{figure}

To efficiently approach the GM point on PES, we integrated the MEALPY library$^{\text{\small\cite{mealpy}}}$ into HASGO. The MEALPY library was also used for the other benchmark nature-inspired metaheuristic GO algorithms. Fig.~\ref{Sfig:HS}a illustrates the operating principle of the Harmony Search (HS) algorithm in HASGO. The performance and convergence of HS rely on the judicious selection of its core hyperparameters.  The total computational budget is defined by the Epoch parameter, which specifies the total number of search iterations performed. The population size determines the number of candidate solutions (harmonies) maintained and processed within the Harmony Memory in each generation, directly impacting the diversity of the explored space. The algorithm's search dynamics are primarily controlled by parameters governing the Pitch Adjustment operation, which uses the Bandwidth (${BW}$). The Forward Parameter (${fw\_para}$) acts as the initial scaling factor, determining the starting magnitude of the ${BW}$ relative to the entire search domain, thereby setting the initial extent of global exploration. Subsequently, the Forward Damping (${fw\_damp}$) factor controls the dynamic reduction of this ${BW}$ over successive epochs. By setting ${fw\_damp}$ close to but less than 1.0, the algorithm facilitates a crucial transition from wide-ranging exploration in early stages to precise local exploitation as it converges toward the optimal solution. For all work in this paper, ${fw\_para}$ is set to 0.05, ${fw\_damp}$ is set to 0.999.

\newpage
\section{The Gibbs free energy computation}
\label{sec:Gibbs_free_energy}
In this work, the chemical potentials of metal atoms were referenced to their corresponding bulk phases: $\mu_{\mathrm{Pd}}$, $\mu_{\mathrm{Cu}}$, and $\mu_{\mathrm{Ag}}$ were set to the energies of one Pd, Cu, and Ag atom in bulk palladium, copper, and silver, respectively. The chemical potential of ethylene, $\mu_{\mathrm{C_2H_4}}$, was evaluated from standard thermodynamic relations at the specified temperature and pressure. The Gibbs free energy of adsorbed ethylene, $G_{\mathrm{C_2H_4,ads}}$, includes the adsorption entropy as well as thermal corrections to enthalpy and entropy. Its calculated value ranges from -0.35 to -0.25 eV depending on the local adsorption environment. To reduce the computational cost in the ASOP simulations, $G_{\mathrm{C_2H_4,ads}}$ was approximated as a constant value of -0.30 eV.

\par
For $\mu_{\mathrm{O_2}}$, it can be calculated from the thermodynamics equation as

\begin{equation}
    \mu_{\mathrm{O_2}}(T,p^{\theta})=\mu_{\mathrm{O_2}}(T,p^{\theta}=1 bar)+k_BT\ln(p_{\mathrm{O_2}}/p^{\theta})
\end{equation}

 where $\mu_{\mathrm{O_2}}(T,p^{\theta})$ is the zero reference state of $\mu_{\mathrm{O_2}}$. To avoid the use of DFT energy of spin-polarized O$_2$ molecule, the standard molar enthalpy of formation of $\mathrm{H_2O}$ (at 298.15 K, $\Delta_fH_m(\mathrm{H_2O})$ = -2.506 eV) is then used to determine the $\mu_{\mathrm{O_2}}(T,p^{\theta})$: 
 
\begin{equation}
    \mu_{\mathrm{O_2}}(T,p^{\theta})=(H(\mathrm{H_2O})-H(\mathrm{H_2})-\Delta_fH_m(\mathrm{H_2O}))\times2+\Delta\mu_{\mathrm{O_2}}(T,p^{\theta})
\end{equation}

where $H(\mathrm{H_2O})$ and $H(\mathrm{H_2})$ is the enthalpy of gaseous $\mathrm{H_2O}$ and $\mathrm{H_2}$ on 298.15 K, which can be approximated as the ZPE corrected total energies of a $\mathrm{H_2O}$ and a $\mathrm{H_2}$ isolated molecule. $\Delta\mu_{\mathrm{O_2}}(T,p^{\theta})$ item is the chemical potential variation of 1 bar O$_2$ from 298.15 K (in enthalpy) to T (in free energy).

Notably, DFT-D3 dispersion correction was not applied to the Cu/CuO system, following previous studies showing that Cu/CuO energetics are more accurately described without dispersion corrections.$^{\text{\small\cite{CuO_without_D3}}}$

\newpage
\section{DFT Calculations}
\label{sec:dft}
All DFT calculations in the GNN dataset are performed using the plane-wave VASP package$^{\text{\small\cite{VASP}}}$. The kinetic energy cutoff was set to 450 eV, the electron-ion interaction was represented by the projector-augmented wave (PAW) pseudopotential, and the exchange-correlation functional was evaluated with GGA-PBE$^{\text{\small\cite{GGA-PBE}}}$. The Monkhorst-Pack k-mesh was 25 times the reciprocal lattice vectors (1/25 \AA$^{-1}$) for single-point calculations. For all structures, the energy and force convergence criteria were set to less than $5 \times 10^{-6}$ eV and 0.01~$\mathrm{eV \, \text{\AA}^{-1}}$, respectively. The van der Waals interactions were considered using the DFT-D3 method with zero damping$^{\text{\small\cite{DFTD3_1,DFTD3_2}}}$ during the calculation of the structures in the global optimization trajectory, whereas standard DFT (without D3) was employed for the fine-tuning dataset preparation.

\newpage
\section{Convergence}
\label{sec:Convergence}
The HASGO simulation is stopped when the halting criterion is reached. The halting criterion in HASGO is by default set to 10 further generations if the simulation can not find other better structures for systems $\leq$10 atoms per simulation cell$^{\text{\small\cite{calypso_convergence}}}$.

\newpage
\section{MLIPs Dataset Analysis}
\label{sec:finetune dataset analysis}
\subsection{PdO}
\begin{figure}[H]
  \centering
  \includegraphics[scale=0.42]{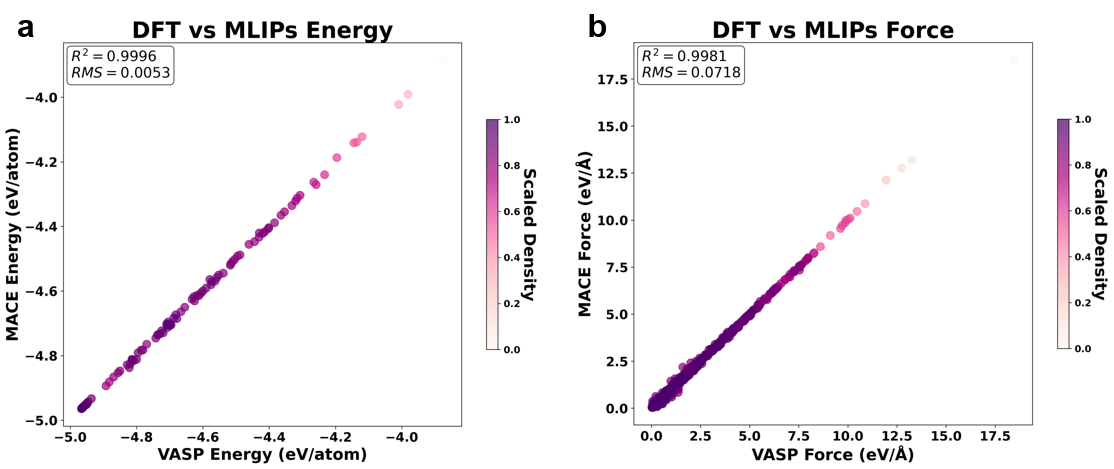}
  \caption{Comparison of DFT reference data and MLIPs predictions for the specified PdO dataset for finetuning with 0.1k data: (Left) energy parity plot and (Right) force parity plot.}
  \label{Sfig:PdO_dft_comparison_0.1k}
\end{figure}

\begin{figure}[H]
  \centering
  \includegraphics[scale=0.42]{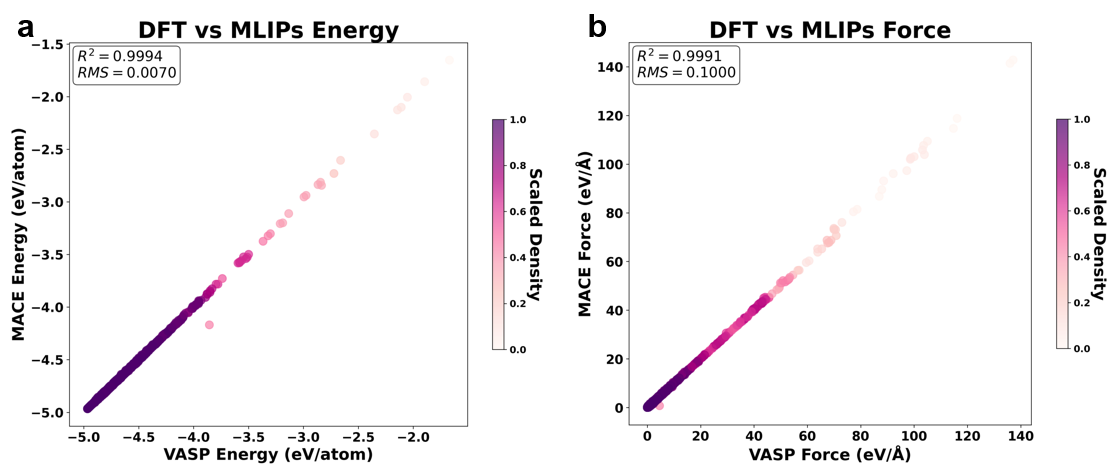}
  \caption{Comparison of DFT reference data and MLIPs predictions for the specified PdO dataset for finetuning with 3k data: (Left) energy parity plot and (Right) force parity plot.}
  \label{Sfig:PdO_dft_comparison_3k}
\end{figure}

\newpage
\subsection{CuO}
\begin{figure}[H]
  \centering
  \includegraphics[scale=0.42]{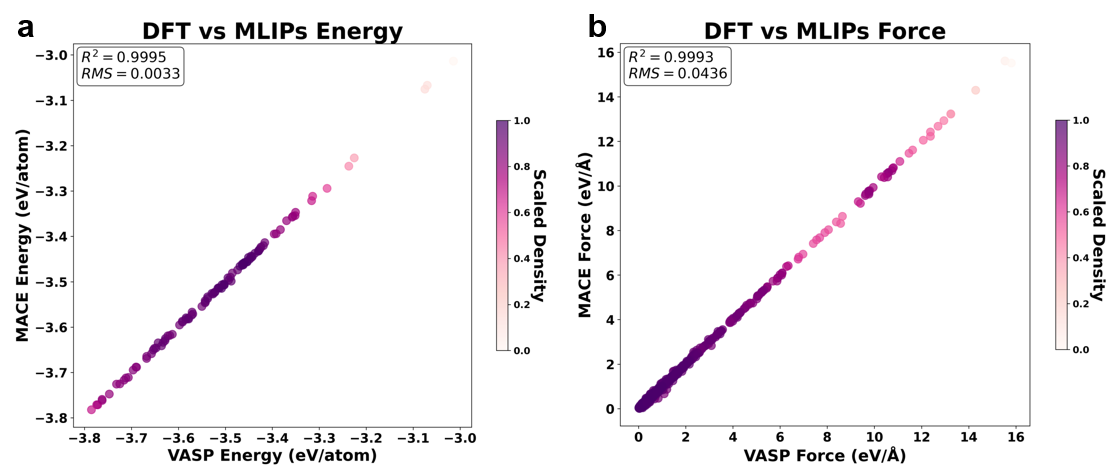}
  \caption{Comparison of DFT reference data and MLIPs predictions for the specified CuO dataset for finetuning with 0.1k data: (Left) energy parity plot and (Right) force parity plot.}
  \label{Sfig:CuO_dft_comparison_0.1k}
\end{figure}

\begin{figure}[H]
  \centering
  \includegraphics[scale=0.42]{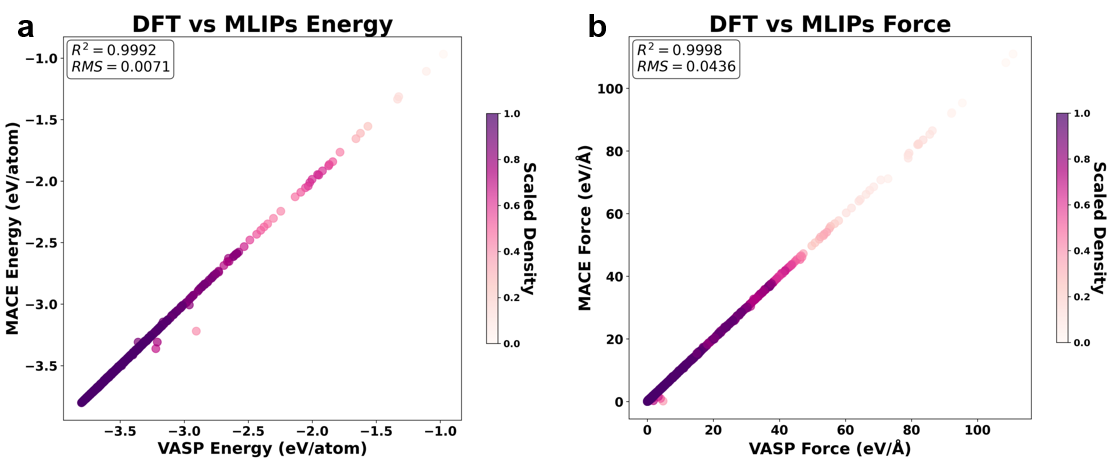}
  \caption{Comparison of DFT reference data and MLIPs predictions for the specified CuO dataset for finetuning with 3k data: (Left) energy parity plot and (Right) force parity plot.}
  \label{Sfig:CuO_dft_comparison_3k}
\end{figure}

\newpage
\subsection{AgO}
\begin{figure}[H]
  \centering
  \includegraphics[scale=0.42]{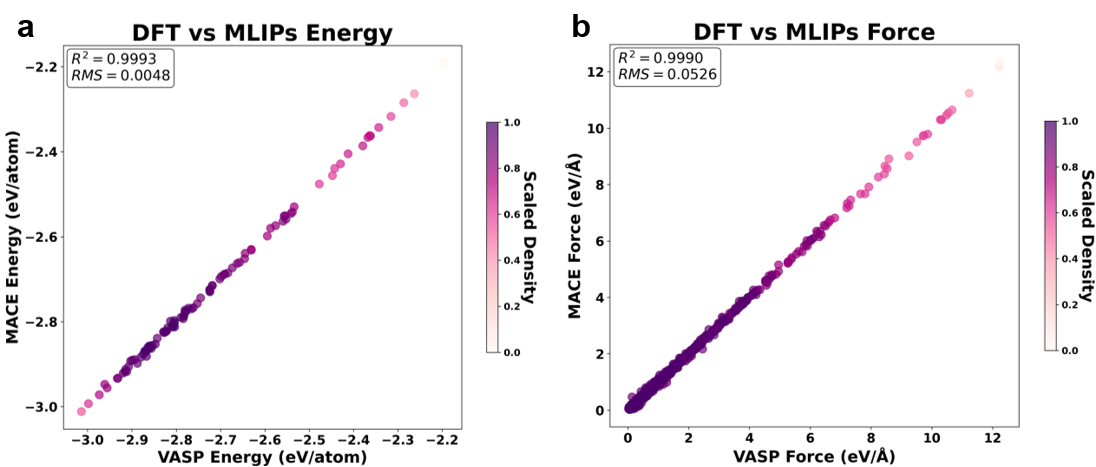}
  \caption{Comparison of DFT reference data and MLIPs predictions for the specified AgO dataset for finetuning with 0.1k data: (Left) energy parity plot and (Right) force parity plot.}
  \label{Sfig:AgO_dft_comparison_0.1k}
\end{figure}

\begin{figure}[H]
  \centering
  \includegraphics[scale=0.42]{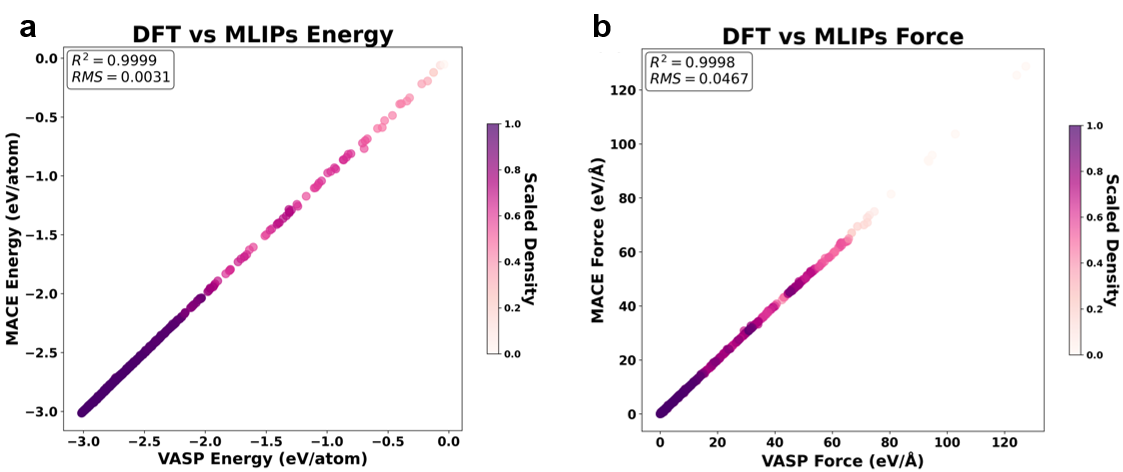}
  \caption{Comparison of DFT reference data and MLIPs predictions for the specified AgO dataset for finetuning with 3k data: (Left) energy parity plot and (Right) force parity plot.}
  \label{Sfig:AgO_dft_comparison_3k}
\end{figure}

\newpage
\subsection{AgCHO}
\begin{figure}[H]
  \centering
  \includegraphics[scale=0.45]{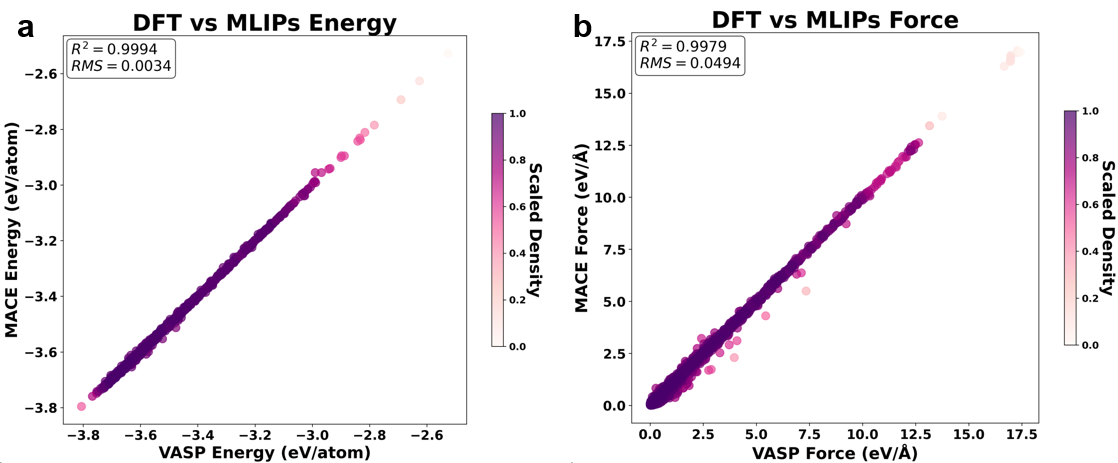}
  \caption{Comparison of DFT reference data and MLIPs predictions for the specified AgCHO dataset for finetuning with 3k data: (Left) energy parity plot and (Right) force parity plot.}
  \label{Sfig:AgCHO_dft_comparison_3k}
\end{figure}

\newpage
\section{Data Coverage}
\label{sec:data coverage}
\begin{figure}[H]
  \centering
  \includegraphics[scale=0.55]{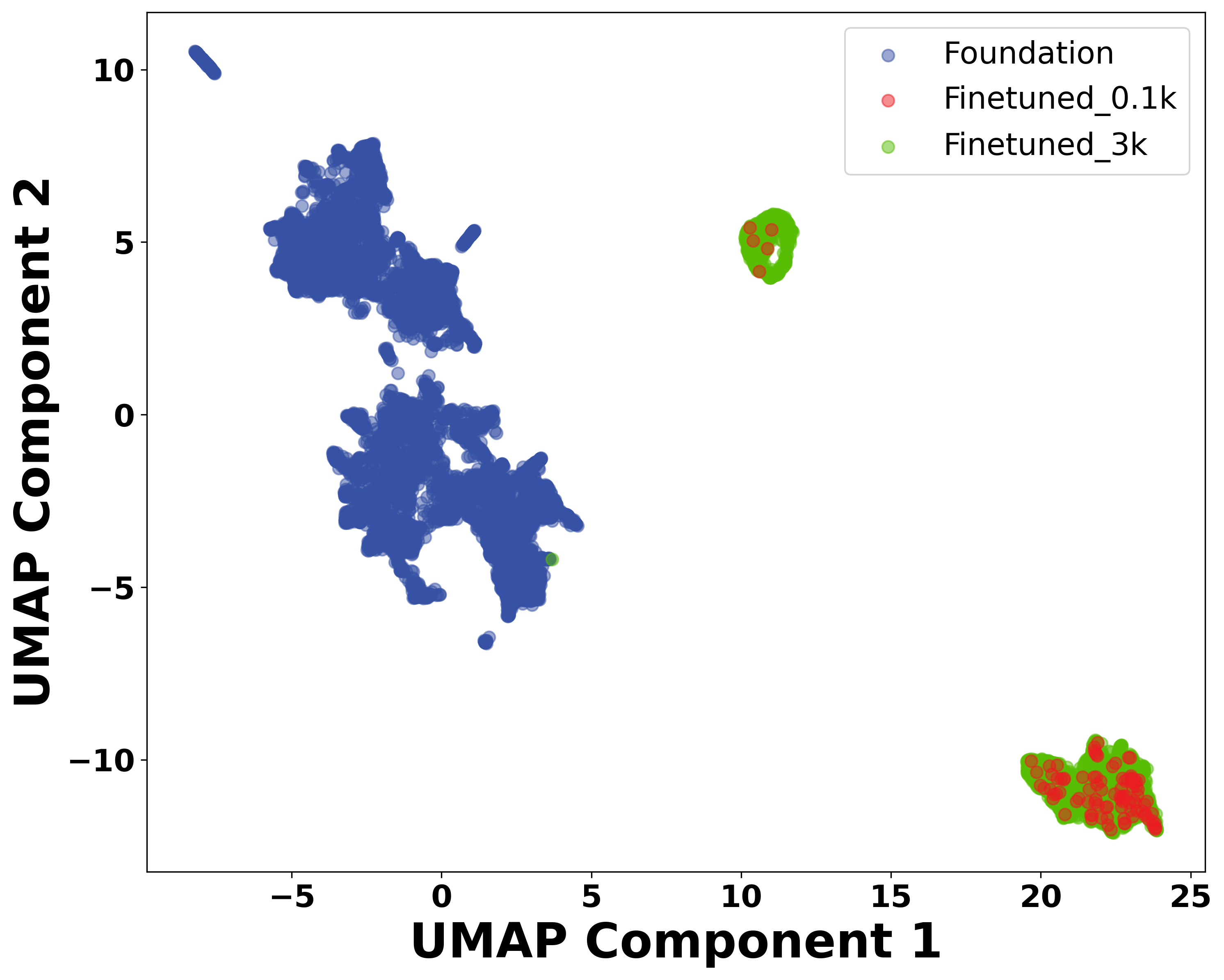}
  \caption{UMAP visualization of chemical environment feature spaces. Two-dimensional projection of local atomic environment descriptors for the PdO-specified dataset and its corresponding replay dataset. The descriptors were extracted from the feature space of the MACE-OMAT foundation model to ensure consistent feature representation. The visualization illustrates the distribution of the 10k samples in the Foundation dataset (blue circles) relative to the Finetuned datasets. Specifically, Finetuned\_0.1k (red circles) represents the distribution for 100 samples for fine-tuning, while Finetuned\_3k (green circles) represents the distribution for 3k samples for fine-tuning.}
  \label{Sfig:PdO_umap}
\end{figure}

\begin{figure}[H]
  \centering
  \includegraphics[scale=0.55]{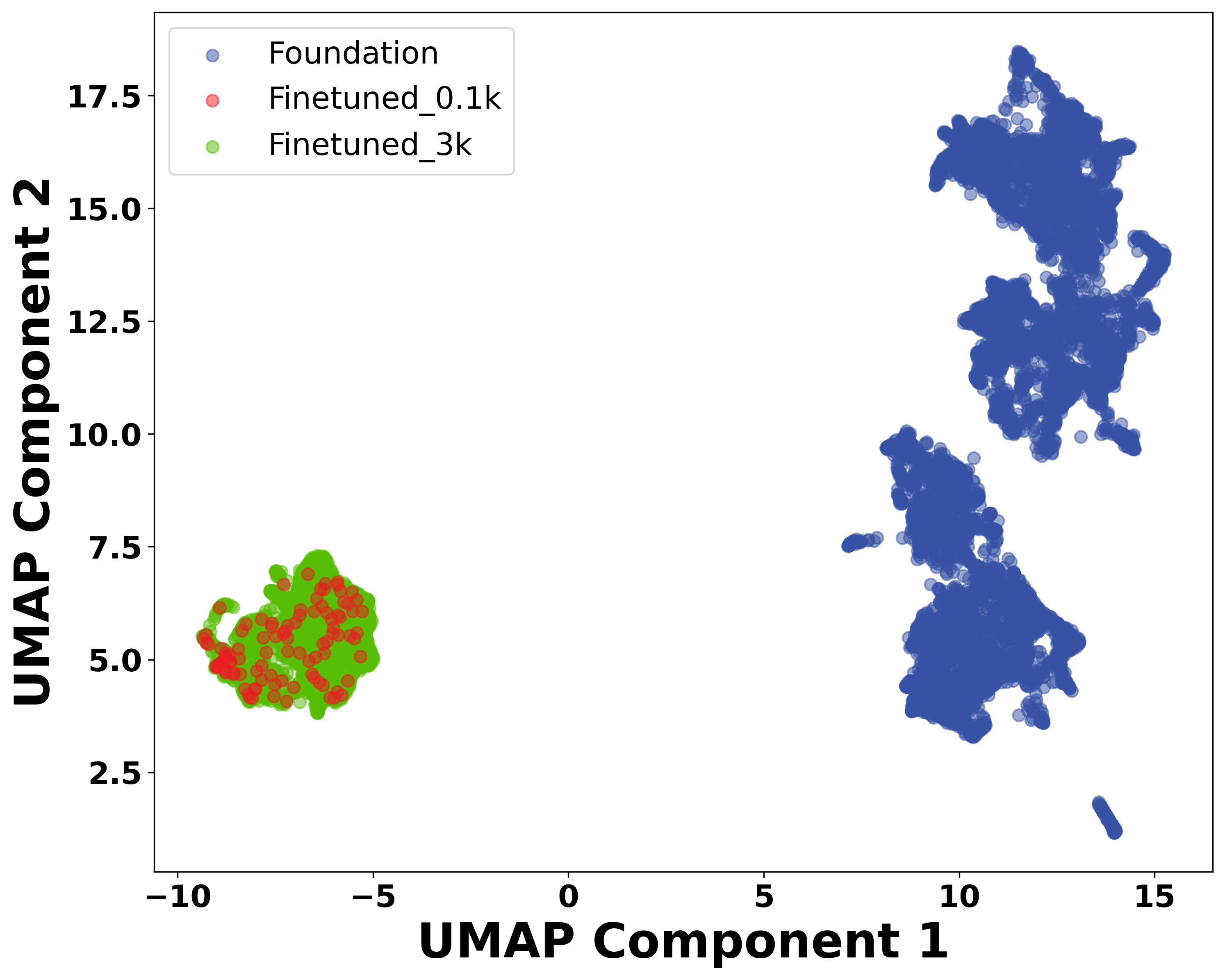}
  \caption{\textbf{UMAP visualization of chemical environment feature spaces.} Two-dimensional projection of local atomic environment descriptors for the CuO-specified dataset and its corresponding replay dataset. The descriptors were extracted from the feature space of the MACE-OMAT foundation model to ensure consistent feature representation. The visualization illustrates the distribution of the 10k samples in the Foundation dataset (blue circles) relative to the Finetuned datasets. Specifically, Finetuned\_0.1k (red circles) represents the distribution for 100 samples for fine-tuning, while Finetuned\_3k (green circles) represents the distribution for 3k samples for fine-tuning.}
  \label{Sfig:CuO_umap}
\end{figure}

\begin{figure}[H]
  \centering
  \includegraphics[scale=0.55]{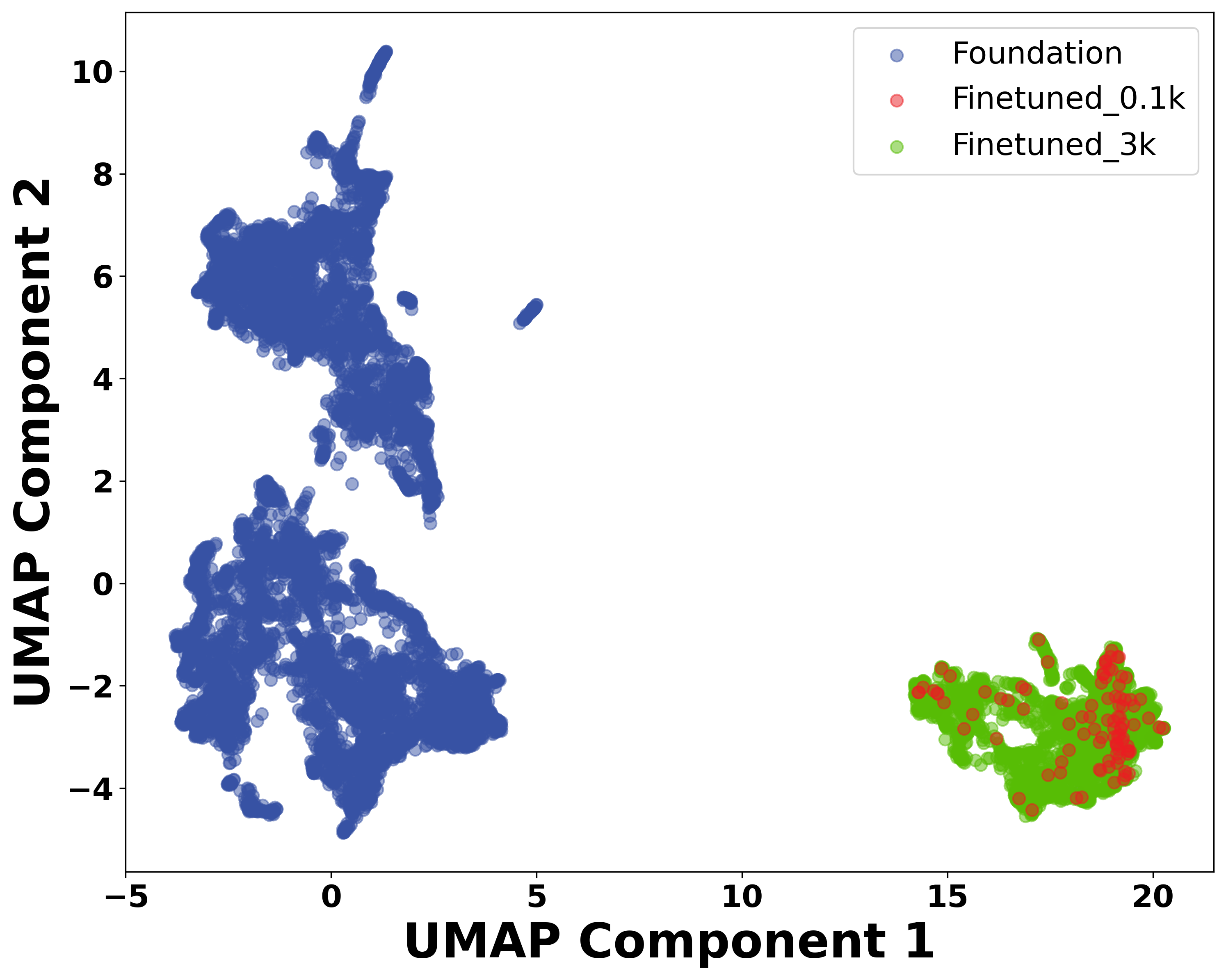}
  \caption{\textbf{UMAP visualization of chemical environment feature spaces.} Two-dimensional projection of local atomic environment descriptors for the AgO-specified dataset and its corresponding replay dataset. The descriptors were extracted from the feature space of the MACE-OMAT foundation model to ensure consistent feature representation. The visualization illustrates the distribution of the 10k samples in Foundation dataset (blue circles) relative to the Finetuned datasets. Specifically, Finetuned\_0.1k (red circles) represents the distribution for 100 samples for fine-tuning, while Finetuned\_3k (green circles) represents the distribution for 3k samples for fine-tuning.}
  \label{Sfig:AgO_umap}
\end{figure}

\begin{figure}[H]
  \centering
  \includegraphics[scale=0.55]{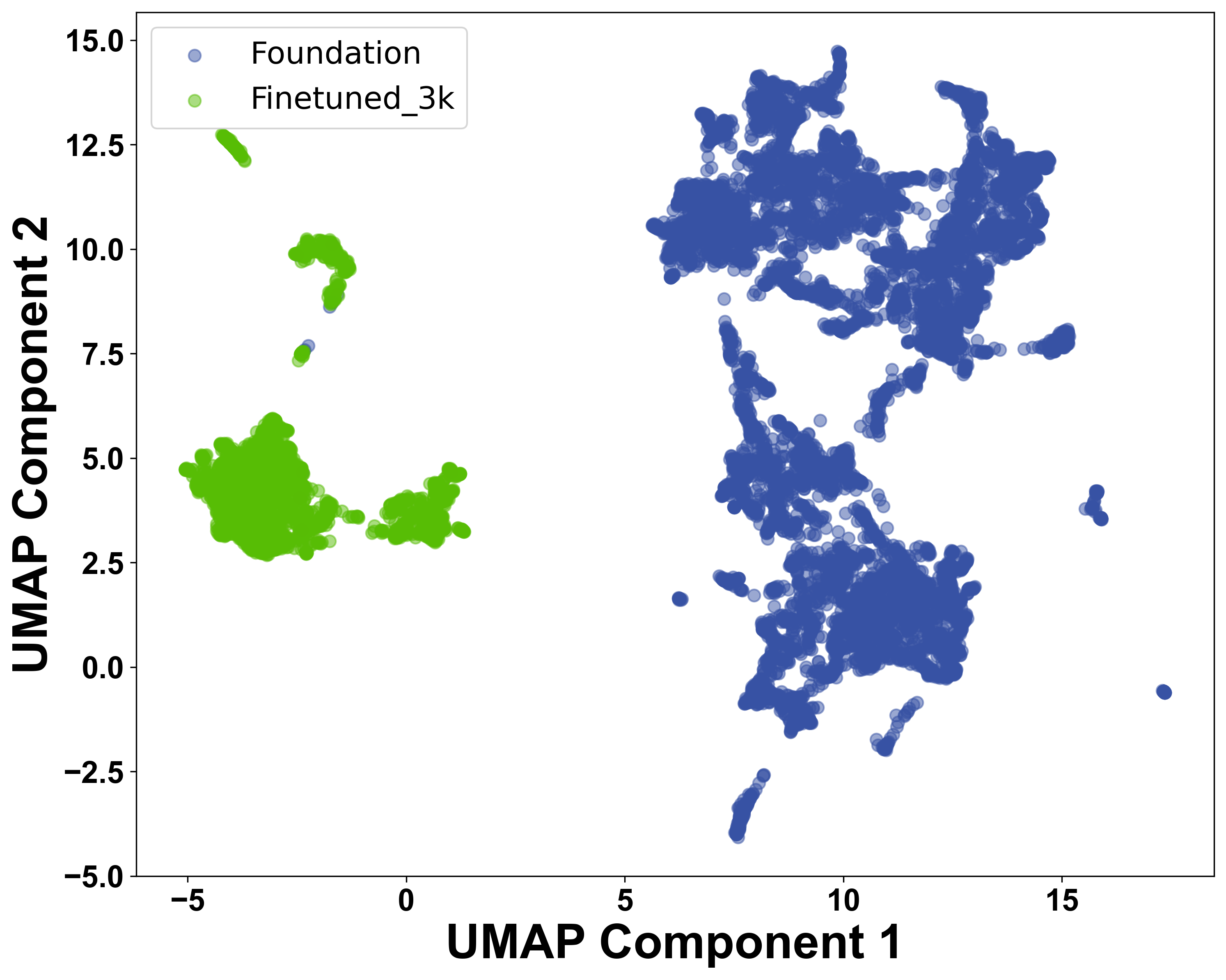}
  \caption{\textbf{UMAP visualization of chemical environment feature spaces.} Two-dimensional projection of local atomic environment descriptors for the AgCHO-specified dataset and its corresponding replay dataset. The descriptors were extracted from the feature space of the MACE-OMAT foundation model to ensure consistent feature representation. The visualization illustrates the distribution of the 10k samples in Foundation dataset (blue circles) relative to the Finetuned datasets with 3k samples (green circles).}
  \label{Sfig:AgCHO_umap}
\end{figure}

After the finetuned datasets finally established, we utilized Uniform Manifold Approximation and Projection (UMAP$^{\text{\small\cite{umap}}}$) to visualize the two-dimensional distribution of local atomic environments. Using descriptors derived from the MACE-OMAT foundation model, we compared the Finetuned\_0.1k, Finetuned\_3k (Specified Dataset in Fig.~\ref{fig:hasgo}b), and Foundation Dataset (Replay Dataset in Fig.~\ref{fig:hasgo}b ensembles. For four surface reconstruction systems, PdO (Supplementary Fig.~\ref{Sfig:PdO_umap}), CuO (Supplementary Fig.~\ref{Sfig:CuO_umap}, AgO (Supplementary Fig.~\ref{Sfig:AgO_umap}) and AgO with adsorbed ethylene (Supplementary Fig.~\ref{Sfig:AgCHO_umap}), the fine-tuned samples reside in distinct, localized clusters separate from the Replay Dataset distribution of the MACE-OMAT foundation model. This clear spatial segregation demonstrates that the newly sampled environments represent OOD data, capturing chemical spaces not previously covered by the foundation model.

\newpage

\newpage
\section{Detailed ASOPs Information}

\begin{figure}[H]
  \centering
  \includegraphics[scale=0.37]{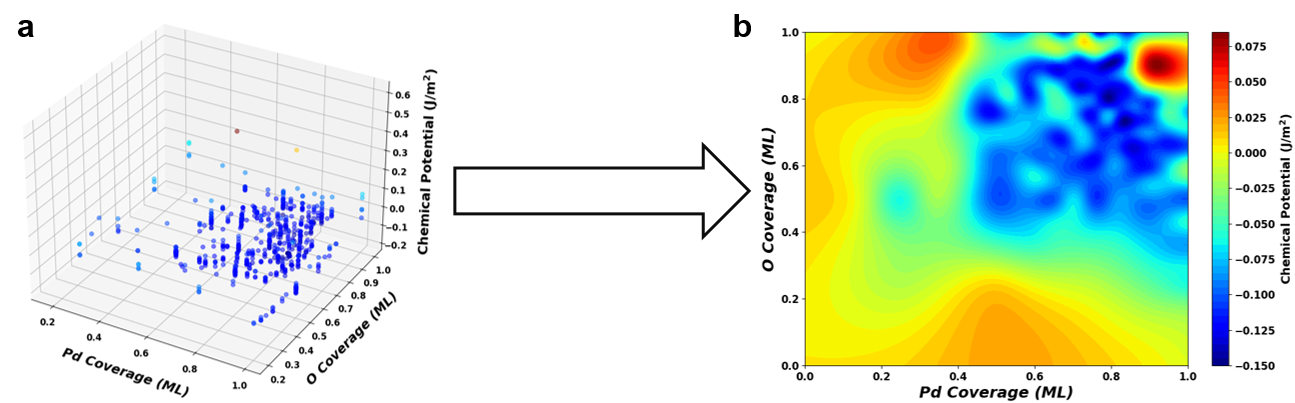}
  \caption{The construction of the PES for Pd(100) surface oxides: \textbf{a.} raw data of $\gamma$ for various compositions undergo minimum projection to  \textbf{b.} the final continuous PES contour plot.}
  \label{sfig:PdO-asop}
\end{figure}

\begin{figure}[H]
  \centering
  \includegraphics[scale=0.37]{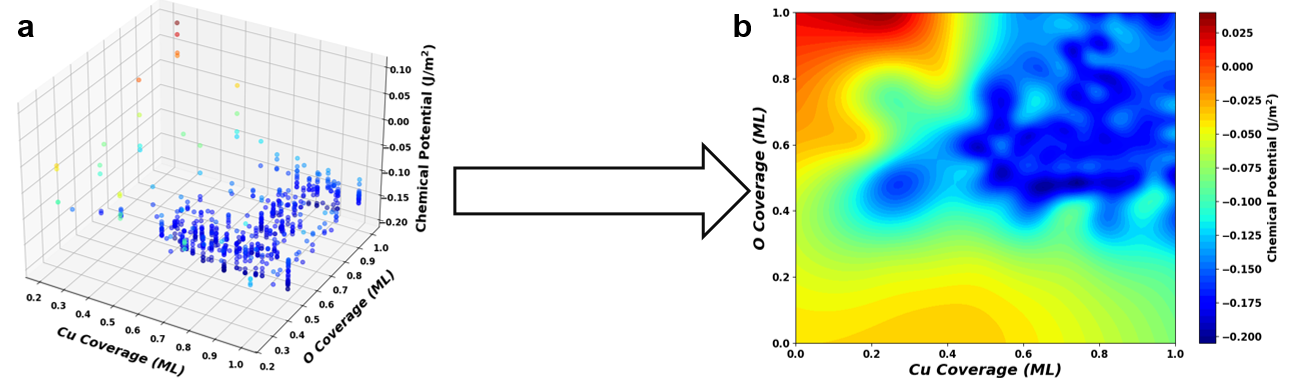}
  \caption{The construction of the PES for Cu(100) surface oxides: \textbf{a.} raw data of $\gamma$ for various compositions undergo minimum projection to  \textbf{b.} the final continuous PES contour plot.}
  \label{sfig:CuO-asop}
\end{figure}

\begin{figure}[H]
  \centering
  \includegraphics[scale=0.37]{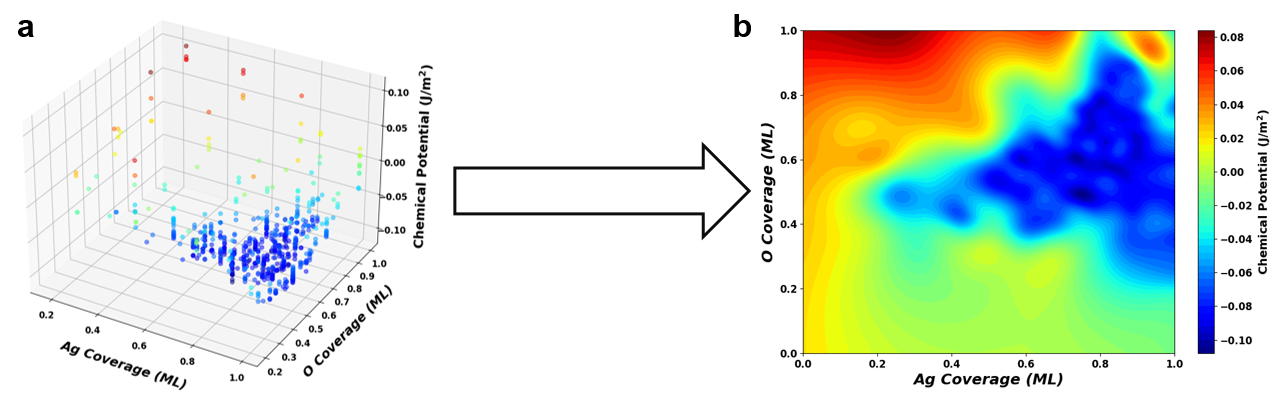}
  \caption{The construction of the PES for Ag(100) surface oxides: \textbf{a.} raw data of $\gamma$ for various compositions undergo minimum projection to  \textbf{b.} the final continuous PES contour plot.}
  \label{sfig:AgO-asop}
\end{figure}

\begin{figure}[H]
  \centering
  \includegraphics[scale=0.37]{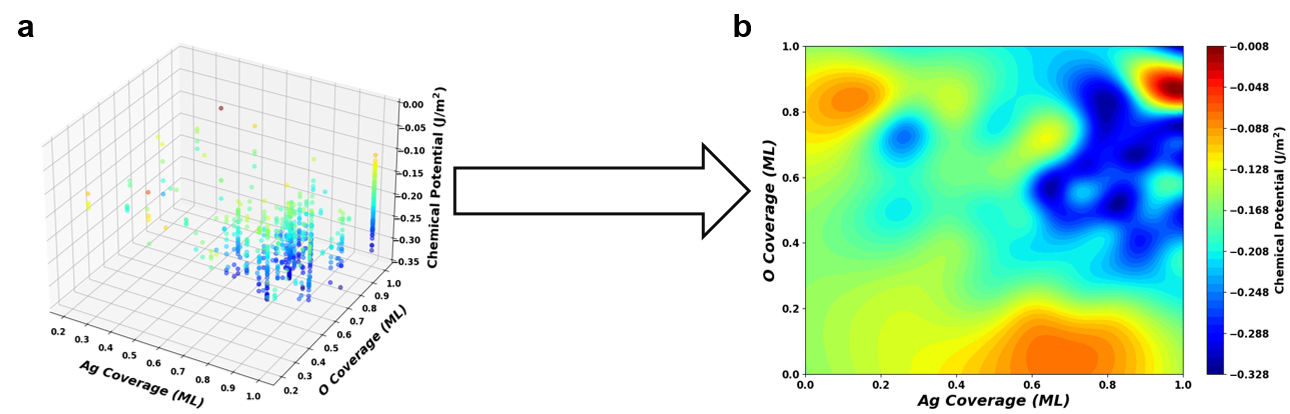}
  \caption{The construction of the PES for Ag(100) surface oxides with adsorbed ethylene: \textbf{a.} raw data of $\gamma$ for various compositions undergo minimum projection to  \textbf{b.} the final continuous PES contour plot.}
  \label{sfig:AgO-ethene-asop}
\end{figure}

\newpage
\section{Surface Reconstruction within High-Symmetry Unit Cells}

\begin{figure}[H]
  \centering
  \includegraphics[scale=0.45]{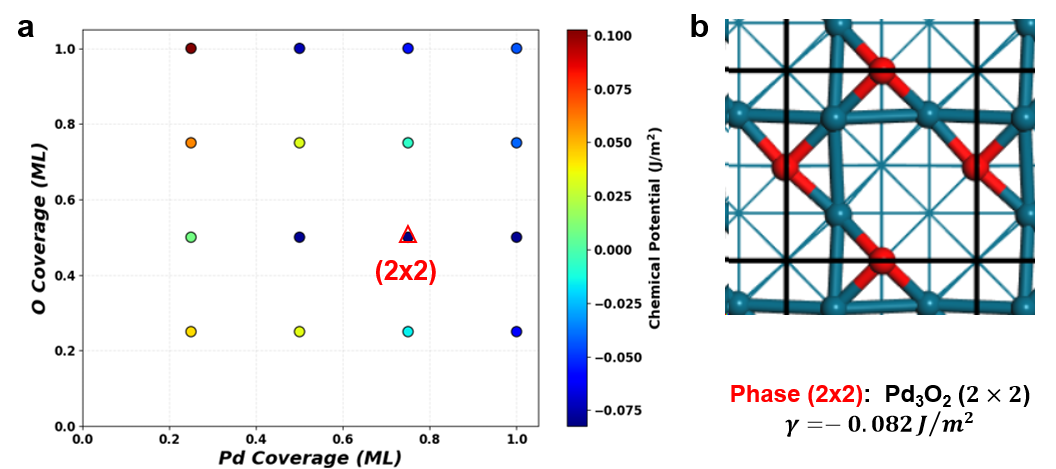}
  \caption{\textbf{The metal-oxides surface reconstruction simulation on Pd(100) under STM imaging condition at $T = 573$~K and $P_{\mathrm{O_2}} = 5 \times 10^{-5}$~mbar.} \textbf{a.} The PES scatter map for palladium surface oxides on a high-symmetry $p(2\times2)$ Pd(100) unit cell within a GC ensemble. \textbf{b.} The $(2\times2)$ phase, identified as the global minimum under this condition within the high-symmetry $p(2\times2)$ unit cell.}
  \label{Sfig:Pd100-high-symmetry}
\end{figure}

\begin{figure}[H]
  \centering
  \includegraphics[scale=0.45]{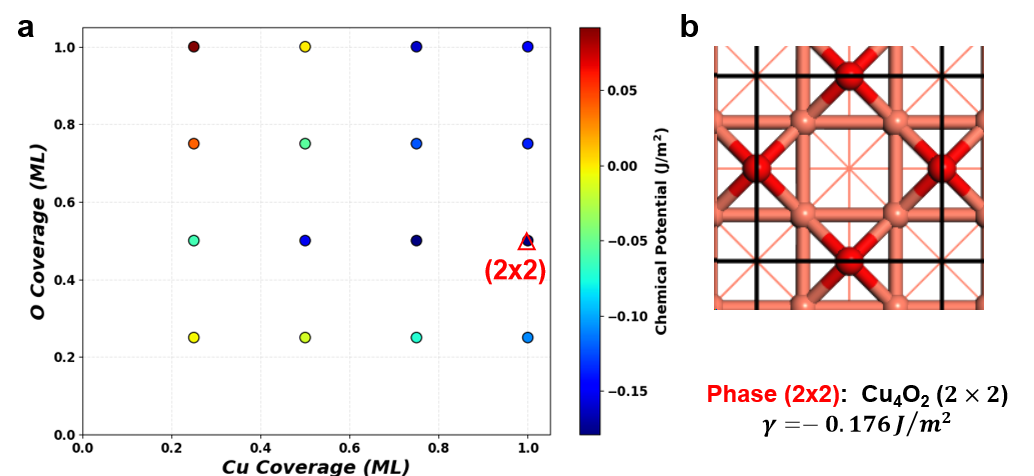}
  \caption{\textbf{The metal-oxides surface reconstruction simulation on Cu(100) under STM imaging condition at $T = 673$~K and $P_{\mathrm{O_2}} = 7 \times 10^{-5}$~mbar.} \textbf{a.} The PES scatter map for copper surface oxides on a high-symmetry $p(2\times2)$ Cu(100) unit cell within a GC ensemble. \textbf{b.} The $(2\times2)$ phase, identified as the global minimum under this condition within the high-symmetry $p(2\times2)$ unit cell.}
  \label{Sfig:Cu100-high-symmetry}
\end{figure}

\begin{figure}[H]
  \centering
  \includegraphics[scale=0.45]{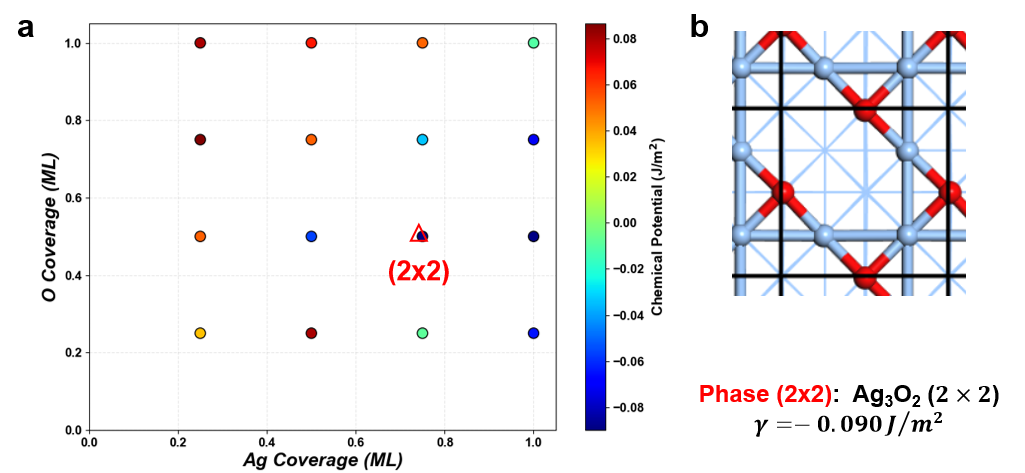}
  \caption{\textbf{The metal-oxides surface reconstruction simulation on Ag(100) under STM imaging condition at $T = 440$~K and $P_{\mathrm{O_2}} = 1 \times 10^{-1}$~mbar.} \textbf{a.} The PES scatter map for silver surface oxides on a high-symmetry $p(2\times2)$ Ag(100) unit cell within a GC ensemble. \textbf{b.} The $(2\times2)$ phase, identified as the global minimum under this condition within the high-symmetry $p(2\times2)$ unit cell.}
  \label{Sfig:Ag100-high-symmetry}
\end{figure}

\begin{figure}[H]
  \centering
  \includegraphics[scale=0.45]{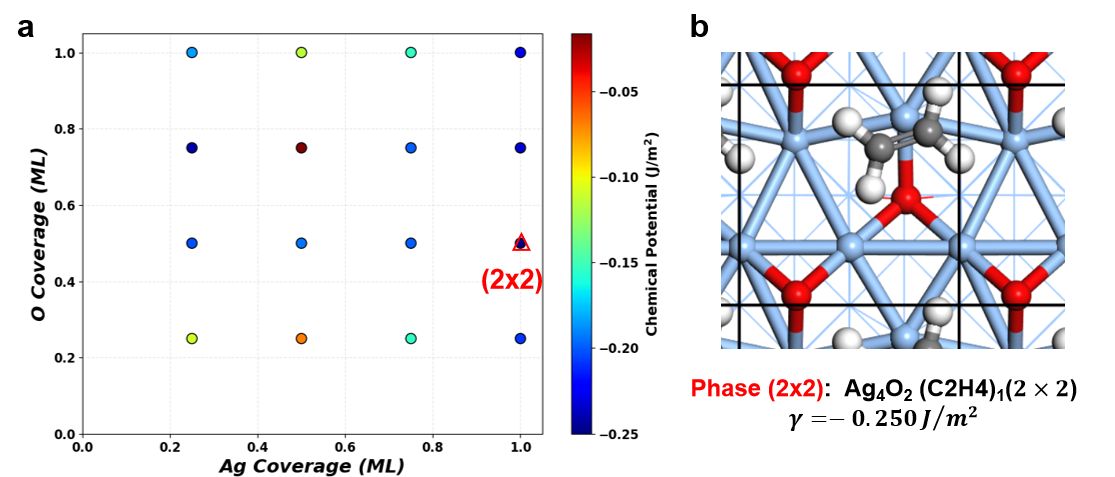}
  \caption{\textbf{The metal-oxides surface reconstruction simulation on Ag(100) under ethylene epoxidation at $T = 500$~K, $P_{\mathrm{O_2}} = 1 $~bar, and $P_{\mathrm{C_2H_4}} = 1 $~bar.} \textbf{a.} The PES scatter map for silver surface oxides on a high-symmetry $p(2\times2)$ Ag(100) unit cell within a GC ensemble. \textbf{b.} The $(2\times2)$ phase, identified as the global minimum under this condition within the high-symmetry $p(2\times2)$ unit cell.}
  \label{Sfig:Ag100-ethene}
\end{figure}

\newpage
\section{Surface Reconstruction under Operating Conditions}
\label{sec:operating_conditions}

\begin{figure}[H]
  \centering
  \includegraphics[scale=0.35]{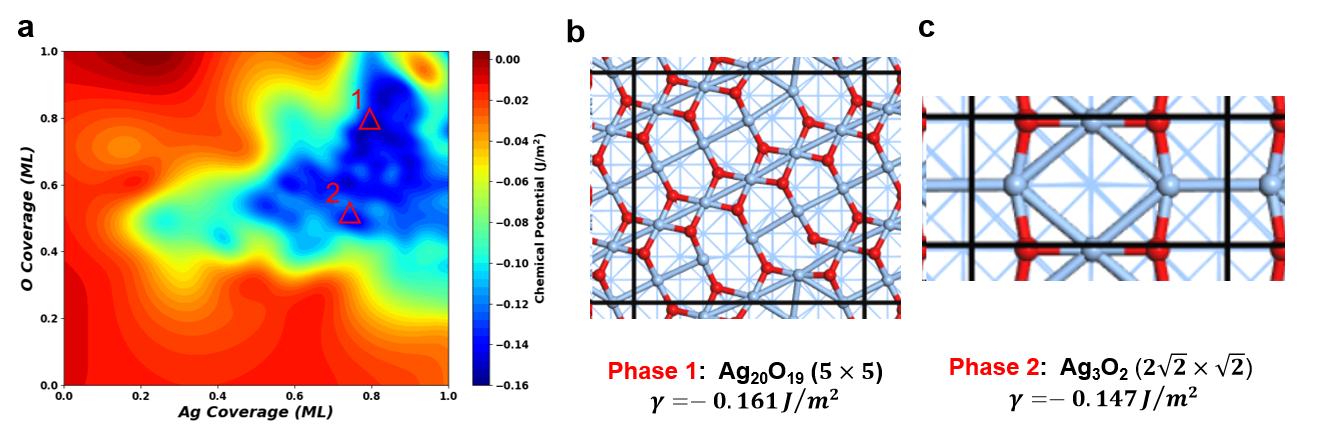}
  \caption{\textbf{The metal-oxides surface reconstruction simulation on Ag(100) under industrial operating condition ($T = 500$~K, $P_{\mathrm{O_2}} = 1$~bar).} \textbf{a.} The PES contour map for silver surface oxides on Ag(100) from ASOPs simulation. \textbf{b.} Phase 1 (Ag$_{20}$O$_{19}$, $(5 \times 5)$, $\gamma = -0.161~\mathrm{J \cdot \text{m}^{-2}}$), which can be characterized as the Ag(101) overlayer structure with one missing oxygen atoms, leading to higher oxygen coverage. \textbf{c.} Phase 2 (Ag$_{3}$O$_{2}$, $(2\sqrt{2} \times \sqrt{2})$ $R45^\circ$, $\gamma = -0.147~\mathrm{J \cdot \text{m}^{-2}}$), which corresponds to the most stable phase identified under high-vacuum condition (Phase 1 in Fig.~\ref{fig:Ag100}b). Phase 1 in this figure is the global minimum under this specific operating condition.}
  \label{Sfig:AgO-operating}
\end{figure}

\newpage
\section{ASOP Explored Supercell Matrices and Compositions}
\subsection{Pd(100)}

Table~\ref{tab:asop_pd} lists all symmetry-inequivalent supercell transformation
matrices $\mathbf{H}$ grouped by $A = |\det\mathbf{H}|$, together with the
specific $(N_\mathrm{Pd}, N_\mathrm{O})$ compositions explored under each.

\begin{longtable}{ccl}
  \caption{Supercell vectors $[u, v]$ and their explored compositions $(N_\mathrm{Pd}, N_\mathrm{O})$ in ASOP for Pd(100).}
  \label{tab:asop_pd} \\
  \hline
  $A$ & $[u, v]$ & \multicolumn{1}{c}{Explored $(N_\mathrm{Pd}, N_\mathrm{O})$} \\
  \hline
  \endfirsthead
  \hline
  $A$ & $[u, v]$ & \multicolumn{1}{c}{Explored $(N_\mathrm{Pd}, N_\mathrm{O})$} \\
  \hline
  \endhead
    \multirow{4}{*}{$4$} & $\left[\begin{smallmatrix}1 & 3\\-1 & 1\end{smallmatrix}\right]$ & \makecell[l]{(1,1), (1,2), (1,3), (1,4), (2,1), (2,2), (2,3), (2,4) \\ (3,1), (3,2), (3,3), (3,4), (4,1), (4,2), (4,3), (4,4)} \\
    & $\left[\begin{smallmatrix}2 & -2\\1 & 1\end{smallmatrix}\right]$ & \makecell[l]{(1,1), (1,2), (1,3), (1,4), (2,1), (2,2), (2,3), (2,4) \\ (3,1), (3,2), (3,3), (3,4), (4,1), (4,2), (4,3), (4,4)} \\ \cline{2-3}
    & $\left[\begin{smallmatrix}2 & 0\\0 & 2\end{smallmatrix}\right]$ & \makecell[l]{(1,1), (1,2), (1,3), (1,4), (2,1), (2,2), (2,3), (2,4) \\ (3,1), (3,2), (3,3), (3,4), (4,1), (4,2), (4,3), (4,4)} \\ \cline{2-3}
    & $\left[\begin{smallmatrix}2 & 1\\0 & 2\end{smallmatrix}\right]$ & \makecell[l]{(1,1), (1,2), (1,3), (1,4), (2,1), (2,2), (2,3), (2,4) \\ (3,1), (3,2), (3,3), (3,4), (4,1), (4,2), (4,3), (4,4)} \\ \hline
    \multirow{2}{*}{$5$} & $\left[\begin{smallmatrix}1 & -2\\2 & 1\end{smallmatrix}\right]$ & \makecell[l]{(1,2), (2,2), (2,3), (3,2), (3,3), (3,4), (3,5), (4,2) \\ (4,3), (4,4), (4,5), (5,1), (5,2), (5,3), (5,4)} \\
    & $\left[\begin{smallmatrix}2 & 3\\-1 & 1\end{smallmatrix}\right]$ & \makecell[l]{(2,2), (2,4), (2,5), (3,2), (3,3), (3,4), (3,5), (4,2) \\ (4,3), (4,4), (4,5), (5,1), (5,2), (5,4), (5,5)} \\ \hline
    \multirow{4}{*}{$6$} & $\left[\begin{smallmatrix}2 & -2\\2 & 1\end{smallmatrix}\right]$ & \makecell[l]{(2,3), (3,3), (3,4), (3,5), (3,6), (4,3), (4,4), (4,5) \\ (4,6), (5,3), (5,4), (5,5), (5,6), (6,2), (6,3)} \\
    & $\left[\begin{smallmatrix}3 & 0\\0 & 2\end{smallmatrix}\right]$ & \makecell[l]{(1,3), (3,3), (3,4), (3,5), (3,6), (4,3), (4,4), (4,5) \\ (4,6), (5,3), (5,4), (5,5), (5,6), (6,2), (6,3)} \\ \cline{2-3}
    & $\left[\begin{smallmatrix}3 & 0\\1 & 2\end{smallmatrix}\right]$ & \makecell[l]{(3,3), (3,4), (3,5), (3,6), (4,3), (4,4), (4,5), (4,6) \\ (5,3), (5,4), (5,5), (5,6), (6,2), (6,3), (6,4)} \\ \cline{2-3}
    & $\left[\begin{smallmatrix}3 & 1\\0 & 2\end{smallmatrix}\right]$ & \makecell[l]{(3,3), (3,4), (3,5), (3,6), (4,3), (4,4), (4,5), (4,6) \\ (5,3), (5,4), (5,5), (5,6), (6,2), (6,3), (6,4)} \\ \hline
    \multirow{1}{*}{$7$} & $\left[\begin{smallmatrix}1 & 3\\-2 & 1\end{smallmatrix}\right]$ & \makecell[l]{(4,3), (4,4), (4,5), (4,6), (5,3), (5,4), (5,5), (5,6) \\ (5,7), (6,4), (6,5), (6,7), (7,2), (7,3), (7,5)} \\ \hline
    \multirow{7}{*}{$8$} & $\left[\begin{smallmatrix}1 & 3\\-2 & 2\end{smallmatrix}\right]$ & \makecell[l]{(4,4), (4,6), (5,4), (5,5), (5,7), (6,4), (6,5), (6,6) \\ (6,7), (6,8), (7,4), (7,5), (7,6), (7,8), (8,4)} \\
    & $\left[\begin{smallmatrix}2 & -3\\2 & 1\end{smallmatrix}\right]$ & \makecell[l]{(4,4), (4,6), (5,4), (5,5), (5,6), (6,4), (6,5), (6,6) \\ (6,7), (6,8), (7,4), (7,5), (7,6), (7,8), (8,4)} \\ \cline{2-3}
    & $\left[\begin{smallmatrix}2 & -2\\2 & 2\end{smallmatrix}\right]$ & \makecell[l]{(4,4), (4,6), (5,4), (5,5), (5,7), (6,4), (6,5), (6,6) \\ (6,7), (6,8), (7,4), (7,5), (7,6), (7,8), (8,4)} \\ \cline{2-3}
    & $\left[\begin{smallmatrix}4 & 0\\0 & 2\end{smallmatrix}\right]$ & \makecell[l]{(4,4), (4,6), (5,4), (5,5), (5,7), (6,4), (6,5), (6,6) \\ (6,7), (6,8), (7,4), (7,5), (7,6), (7,8), (8,4)} \\ \cline{2-3}
    & $\left[\begin{smallmatrix}4 & 0\\1 & 2\end{smallmatrix}\right]$ & \makecell[l]{(4,4), (4,6), (5,4), (5,5), (5,7), (6,4), (6,5), (6,6) \\ (6,7), (6,8), (7,4), (7,5), (7,6), (7,8), (8,4)} \\ \cline{2-3}
    & $\left[\begin{smallmatrix}4 & 1\\0 & 2\end{smallmatrix}\right]$ & \makecell[l]{(4,4), (4,6), (5,4), (5,5), (5,7), (6,4), (6,5), (6,6) \\ (6,7), (6,8), (7,4), (7,5), (7,6), (7,8), (8,4)} \\ \cline{2-3}
    & $\left[\begin{smallmatrix}4 & 2\\0 & 2\end{smallmatrix}\right]$ & \makecell[l]{(4,4), (4,6), (5,4), (5,5), (5,7), (6,4), (6,5), (6,6) \\ (6,7), (6,8), (7,4), (7,5), (7,6), (7,8), (8,4)} \\ \hline
    \multirow{4}{*}{$9$} & $\left[\begin{smallmatrix}1 & 4\\-2 & 1\end{smallmatrix}\right]$ & \makecell[l]{(5,4), (5,7), (5,8), (6,5), (6,6), (6,8), (6,9), (7,5) \\ (7,6), (7,7), (7,8), (7,9), (8,5), (8,6), (8,7)} \\
    & $\left[\begin{smallmatrix}3 & -3\\2 & 1\end{smallmatrix}\right]$ & \makecell[l]{(5,7), (5,8), (6,5), (6,6), (6,8), (6,9), (7,5), (7,6) \\ (7,7), (7,8), (7,9), (8,5), (8,6), (8,7), (9,6)} \\ \cline{2-3}
    & $\left[\begin{smallmatrix}3 & 0\\0 & 3\end{smallmatrix}\right]$ & \makecell[l]{(5,7), (5,8), (6,5), (6,6), (6,8), (6,9), (7,5), (7,6) \\ (7,7), (7,8), (7,9), (8,5), (8,6), (8,7), (9,6)} \\ \cline{2-3}
    & $\left[\begin{smallmatrix}3 & 1\\0 & 3\end{smallmatrix}\right]$ & \makecell[l]{(5,7), (5,8), (6,5), (6,6), (6,8), (6,9), (7,5), (7,6) \\ (7,7), (7,8), (7,9), (8,5), (8,6), (8,7), (9,6)} \\ \hline
    \multirow{6}{*}{$10$} & $\left[\begin{smallmatrix}1 & -3\\3 & 1\end{smallmatrix}\right]$ & \makecell[l]{(5,5), (5,7), (6,5), (6,8), (6,9), (7,6), (7,9), (7,10) \\ (8,7), (8,8), (8,9), (8,10), (9,6), (9,7), (9,8)} \\
    & $\left[\begin{smallmatrix}2 & -4\\2 & 1\end{smallmatrix}\right]$ & \makecell[l]{(5,5), (6,5), (6,8), (6,9), (7,5), (7,6), (7,9), (7,10) \\ (8,6), (8,7), (8,8), (8,9), (8,10), (9,6), (10,5)} \\ \cline{2-3}
    & $\left[\begin{smallmatrix}2 & 3\\-2 & 2\end{smallmatrix}\right]$ & \makecell[l]{(5,5), (6,5), (6,8), (6,9), (7,6), (7,9), (7,10), (8,5) \\ (8,7), (8,8), (8,9), (8,10), (9,6), (9,8), (10,5)} \\ \cline{2-3}
    & $\left[\begin{smallmatrix}5 & 0\\0 & 2\end{smallmatrix}\right]$ & \makecell[l]{(5,5), (6,5), (6,8), (6,9), (7,6), (7,9), (7,10), (8,5) \\ (8,7), (8,8), (8,9), (8,10), (9,6), (9,8), (10,5)} \\ \cline{2-3}
    & $\left[\begin{smallmatrix}5 & 1\\0 & 2\end{smallmatrix}\right]$ & \makecell[l]{(5,5), (6,5), (6,8), (6,9), (7,6), (7,9), (7,10), (8,5) \\ (8,7), (8,8), (8,9), (8,10), (9,6), (9,8), (10,5)} \\ \cline{2-3}
    & $\left[\begin{smallmatrix}5 & 2\\0 & 2\end{smallmatrix}\right]$ & \makecell[l]{(5,5), (6,5), (6,8), (6,9), (7,6), (7,9), (7,10), (8,5) \\ (8,7), (8,8), (8,9), (8,10), (9,6), (9,8), (10,5)} \\ \hline
    \multirow{3}{*}{$11$} & $\left[\begin{smallmatrix}2 & -3\\3 & 1\end{smallmatrix}\right]$ & \makecell[l]{(6,9), (7,7), (7,10), (7,11), (8,6), (8,7), (8,9), (8,10) \\ (8,11), (9,8), (9,9), (9,10), (9,11), (10,7), (10,8)} \\
    & $\left[\begin{smallmatrix}3 & -4\\2 & 1\end{smallmatrix}\right]$ & \makecell[l]{(7,6), (7,7), (7,10), (8,6), (8,7), (8,8), (8,9), (8,10) \\ (9,7), (9,8), (9,9), (9,10), (9,11), (10,7), (10,8)} \\ \cline{2-3}
    & $\left[\begin{smallmatrix}5 & -1\\1 & 2\end{smallmatrix}\right]$ & \makecell[l]{(7,5), (7,7), (7,10), (8,6), (8,7), (8,8), (8,9), (8,10) \\ (9,7), (9,8), (9,9), (9,10), (9,11), (10,7), (10,8)} \\ \hline
    \multirow{11}{*}{$12$} & $\left[\begin{smallmatrix}2 & 3\\-2 & 3\end{smallmatrix}\right]$ & \makecell[l]{(6,6), (6,9), (7,10), (8,8), (8,11), (8,12), (9,6), (9,7) \\ (9,9), (9,10), (9,11), (10,9), (10,11), (10,12), (11,9)} \\
    & $\left[\begin{smallmatrix}2 & 4\\-2 & 2\end{smallmatrix}\right]$ & \makecell[l]{(6,6), (6,9), (7,10), (7,11), (8,8), (8,12), (9,6), (9,7) \\ (9,8), (9,9), (9,10), (10,9), (10,10), (10,11), (10,12)} \\ \cline{2-3}
    & $\left[\begin{smallmatrix}3 & -3\\2 & 2\end{smallmatrix}\right]$ & \makecell[l]{(6,6), (6,9), (7,10), (8,8), (8,12), (9,6), (9,7), (9,9) \\ (9,10), (10,8), (10,9), (10,10), (10,11), (10,12), (11,8)} \\ \cline{2-3}
    & $\left[\begin{smallmatrix}3 & -3\\3 & 1\end{smallmatrix}\right]$ & \makecell[l]{(6,6), (6,9), (7,10), (8,8), (8,12), (9,6), (9,7), (9,9) \\ (9,10), (10,8), (10,9), (10,10), (10,11), (10,12), (11,8)} \\ \cline{2-3}
    & $\left[\begin{smallmatrix}4 & -4\\2 & 1\end{smallmatrix}\right]$ & \makecell[l]{(6,6), (6,9), (7,10), (8,8), (8,12), (9,6), (9,7), (9,9) \\ (9,10), (10,8), (10,9), (10,10), (10,11), (10,12), (11,8)} \\ \cline{2-3}
    & $\left[\begin{smallmatrix}4 & 0\\0 & 3\end{smallmatrix}\right]$ & \makecell[l]{(6,6), (6,9), (7,10), (8,8), (8,12), (9,6), (9,7), (9,9) \\ (9,10), (10,8), (10,9), (10,10), (10,11), (10,12), (11,8)} \\ \cline{2-3}
    & $\left[\begin{smallmatrix}4 & 0\\1 & 3\end{smallmatrix}\right]$ & \makecell[l]{(6,6), (6,9), (7,10), (8,8), (8,12), (9,6), (9,7), (9,9) \\ (9,10), (10,8), (10,9), (10,10), (10,11), (10,12), (11,8)} \\ \cline{2-3}
    & $\left[\begin{smallmatrix}4 & 1\\0 & 3\end{smallmatrix}\right]$ & \makecell[l]{(6,6), (6,9), (7,10), (8,8), (8,12), (9,6), (9,7), (9,9) \\ (9,10), (10,8), (10,9), (10,10), (10,11), (10,12), (11,8)} \\ \cline{2-3}
    & $\left[\begin{smallmatrix}4 & 2\\0 & 3\end{smallmatrix}\right]$ & \makecell[l]{(6,6), (6,9), (7,10), (8,8), (8,12), (9,6), (9,7), (9,9) \\ (9,10), (10,8), (10,9), (10,10), (10,11), (10,12), (11,8)} \\ \cline{2-3}
    & $\left[\begin{smallmatrix}5 & -2\\1 & 2\end{smallmatrix}\right]$ & \makecell[l]{(6,6), (6,9), (7,10), (8,8), (8,12), (9,6), (9,7), (9,9) \\ (9,10), (10,8), (10,9), (10,10), (10,11), (10,12), (11,8)} \\ \cline{2-3}
    & $\left[\begin{smallmatrix}6 & 0\\1 & 2\end{smallmatrix}\right]$ & \makecell[l]{(6,6), (6,9), (7,10), (8,8), (8,12), (9,6), (9,7), (9,9) \\ (9,10), (10,8), (10,9), (10,10), (10,11), (10,12), (11,8)} \\ \hline
    \multirow{1}{*}{$16$} & $\left[\begin{smallmatrix}4 & 0\\0 & 4\end{smallmatrix}\right]$ & \makecell[l]{(10,14), (11,14), (11,16), (12,12), (12,13), (12,14), (13,11), (13,12) \\ (13,13), (13,14), (13,15), (13,16), (14,10), (14,12), (14,13)} \\ \hline
    \multirow{1}{*}{$25$} & $\left[\begin{smallmatrix}5 & 0\\0 & 5\end{smallmatrix}\right]$ & \makecell[l]{(15,21), (17,25), (18,22), (19,20), (19,21), (19,22), (20,18), (20,20) \\ (20,21), (20,22), (21,18), (21,19), (21,20), (21,23), (22,19)} \\ \hline
  \hline
\end{longtable}

\clearpage
\subsection{Cu(100)}

Table~\ref{tab:asop_cu} lists all symmetry-inequivalent supercell transformation
matrices $\mathbf{H}$ grouped by $A = |\det\mathbf{H}|$, together with the
specific $(N_\mathrm{Cu}, N_\mathrm{O})$ compositions explored under each,
without vdW correction.

\begin{longtable}{ccl}
  \caption{Supercell vectors $[u, v]$ and their explored compositions $(N_\mathrm{Cu}, N_\mathrm{O})$ in ASOP for Cu(100) without vdW correction.\label{tab:asop_cu}} \\
  \hline
  $A$ & $[u, v]$ & \multicolumn{1}{c}{Explored $(N_\mathrm{Cu}, N_\mathrm{O})$} \\
  \hline
  \endfirsthead
  \hline
  $A$ & $[u, v]$ & \multicolumn{1}{c}{Explored $(N_\mathrm{Cu}, N_\mathrm{O})$} \\
  \hline
  \endhead
    \multirow{4}{*}{$4$} & $\left[\begin{smallmatrix}1 & 3\\-1 & 1\end{smallmatrix}\right]$ & \makecell[l]{(1,1), (1,2), (1,3), (1,4), (2,1), (2,2), (2,3), (2,4) \\ (3,1), (3,2), (3,3), (3,4), (4,1), (4,2), (4,3), (4,4)} \\
    & $\left[\begin{smallmatrix}2 & -2\\1 & 1\end{smallmatrix}\right]$ & \makecell[l]{(1,1), (1,2), (1,3), (1,4), (2,1), (2,2), (2,3), (2,4) \\ (3,1), (3,2), (3,3), (3,4), (4,1), (4,2), (4,3), (4,4)} \\ \cline{2-3}
    & $\left[\begin{smallmatrix}2 & 0\\0 & 2\end{smallmatrix}\right]$ & \makecell[l]{(1,1), (1,2), (1,3), (1,4), (2,1), (2,2), (2,3), (2,4) \\ (3,1), (3,2), (3,3), (3,4), (4,1), (4,2), (4,3), (4,4)} \\ \cline{2-3}
    & $\left[\begin{smallmatrix}2 & 1\\0 & 2\end{smallmatrix}\right]$ & \makecell[l]{(1,1), (1,2), (1,3), (1,4), (2,1), (2,2), (2,3), (2,4) \\ (3,1), (3,2), (3,3), (3,4), (4,1), (4,2), (4,3), (4,4)} \\ \hline
    \multirow{2}{*}{$5$} & $\left[\begin{smallmatrix}1 & -2\\2 & 1\end{smallmatrix}\right]$ & \makecell[l]{(1,3), (2,3), (2,4), (3,2), (3,3), (3,4), (3,5), (4,2) \\ (4,3), (4,4), (4,5), (5,2), (5,3), (5,4), (5,5)} \\
    & $\left[\begin{smallmatrix}2 & 3\\-1 & 1\end{smallmatrix}\right]$ & \makecell[l]{(1,2), (2,2), (2,3), (3,2), (3,3), (3,4), (3,5), (4,2) \\ (4,3), (4,4), (4,5), (5,2), (5,3), (5,4), (5,5)} \\ \hline
    \multirow{4}{*}{$6$} & $\left[\begin{smallmatrix}2 & -2\\2 & 1\end{smallmatrix}\right]$ & \makecell[l]{(2,3), (3,3), (3,4), (4,3), (4,4), (4,5), (4,6), (5,3) \\ (5,4), (5,5), (5,6), (6,3), (6,4), (6,5), (6,6)} \\
    & $\left[\begin{smallmatrix}3 & 0\\0 & 2\end{smallmatrix}\right]$ & \makecell[l]{(2,3), (3,3), (3,4), (4,3), (4,4), (4,5), (4,6), (5,3) \\ (5,4), (5,5), (5,6), (6,3), (6,4), (6,5), (6,6)} \\ \cline{2-3}
    & $\left[\begin{smallmatrix}3 & 0\\1 & 2\end{smallmatrix}\right]$ & \makecell[l]{(2,3), (3,3), (3,4), (4,3), (4,4), (4,5), (4,6), (5,3) \\ (5,4), (5,5), (5,6), (6,3), (6,4), (6,5), (6,6)} \\ \cline{2-3}
    & $\left[\begin{smallmatrix}3 & 1\\0 & 2\end{smallmatrix}\right]$ & \makecell[l]{(2,3), (3,3), (3,4), (4,3), (4,4), (4,5), (4,6), (5,3) \\ (5,4), (5,5), (5,6), (6,3), (6,4), (6,5), (6,6)} \\ \hline
    \multirow{1}{*}{$7$} & $\left[\begin{smallmatrix}1 & 3\\-2 & 1\end{smallmatrix}\right]$ & \makecell[l]{(4,4), (4,5), (5,3), (5,4), (5,5), (5,6), (5,7), (6,4) \\ (6,5), (6,6), (6,7), (7,3), (7,4), (7,5), (7,7)} \\ \hline
    \multirow{7}{*}{$8$} & $\left[\begin{smallmatrix}1 & 3\\-2 & 2\end{smallmatrix}\right]$ & \makecell[l]{(4,4), (4,5), (5,4), (5,5), (6,4), (6,5), (6,6), (6,7) \\ (6,8), (7,4), (7,5), (7,6), (7,8), (8,4), (8,8)} \\
    & $\left[\begin{smallmatrix}2 & -3\\2 & 1\end{smallmatrix}\right]$ & \makecell[l]{(4,4), (5,4), (5,5), (6,4), (6,5), (6,6), (6,7), (6,8) \\ (7,4), (7,5), (7,6), (7,8), (8,4), (8,5), (8,8)} \\ \cline{2-3}
    & $\left[\begin{smallmatrix}2 & -2\\2 & 2\end{smallmatrix}\right]$ & \makecell[l]{(4,4), (5,4), (5,5), (5,6), (6,4), (6,5), (6,6), (6,7) \\ (6,8), (7,4), (7,5), (7,6), (7,8), (8,4), (8,8)} \\ \cline{2-3}
    & $\left[\begin{smallmatrix}4 & 0\\0 & 2\end{smallmatrix}\right]$ & \makecell[l]{(4,4), (4,5), (5,4), (5,5), (6,4), (6,5), (6,6), (6,7) \\ (6,8), (7,4), (7,5), (7,6), (7,8), (8,4), (8,8)} \\ \cline{2-3}
    & $\left[\begin{smallmatrix}4 & 0\\1 & 2\end{smallmatrix}\right]$ & \makecell[l]{(4,4), (4,5), (5,4), (5,5), (6,4), (6,5), (6,6), (6,7) \\ (6,8), (7,4), (7,5), (7,6), (7,8), (8,4), (8,8)} \\ \cline{2-3}
    & $\left[\begin{smallmatrix}4 & 1\\0 & 2\end{smallmatrix}\right]$ & \makecell[l]{(4,4), (4,5), (5,4), (5,5), (6,4), (6,5), (6,6), (6,7) \\ (6,8), (7,4), (7,5), (7,6), (7,8), (8,4), (8,8)} \\ \cline{2-3}
    & $\left[\begin{smallmatrix}4 & 2\\0 & 2\end{smallmatrix}\right]$ & \makecell[l]{(4,4), (4,5), (5,4), (5,5), (6,4), (6,5), (6,6), (6,7) \\ (6,8), (7,4), (7,5), (7,6), (7,8), (8,4), (8,8)} \\ \hline
    \multirow{4}{*}{$9$} & $\left[\begin{smallmatrix}1 & 4\\-2 & 1\end{smallmatrix}\right]$ & \makecell[l]{(5,5), (5,6), (6,5), (7,5), (7,6), (7,7), (7,8), (7,9) \\ (8,4), (8,5), (8,7), (8,9), (9,4), (9,5), (9,9)} \\
    & $\left[\begin{smallmatrix}3 & -3\\2 & 1\end{smallmatrix}\right]$ & \makecell[l]{(4,5), (5,4), (5,5), (5,6), (6,5), (7,5), (7,6), (7,7) \\ (7,8), (7,9), (8,5), (8,6), (8,7), (8,9), (9,9)} \\ \cline{2-3}
    & $\left[\begin{smallmatrix}3 & 0\\0 & 3\end{smallmatrix}\right]$ & \makecell[l]{(5,5), (5,6), (6,5), (6,6), (7,4), (7,5), (7,6), (7,7) \\ (7,8), (7,9), (8,5), (8,6), (8,7), (8,9), (9,9)} \\ \cline{2-3}
    & $\left[\begin{smallmatrix}3 & 1\\0 & 3\end{smallmatrix}\right]$ & \makecell[l]{(5,5), (5,6), (6,5), (6,6), (7,5), (7,6), (7,7), (7,8) \\ (7,9), (8,5), (8,6), (8,7), (8,9), (9,5), (9,9)} \\ \hline
    \multirow{6}{*}{$10$} & $\left[\begin{smallmatrix}1 & -3\\3 & 1\end{smallmatrix}\right]$ & \makecell[l]{(5,6), (6,5), (6,6), (7,5), (7,6), (8,5), (8,6), (8,7) \\ (8,8), (8,9), (8,10), (9,5), (9,6), (9,10), (10,5)} \\
    & $\left[\begin{smallmatrix}2 & -4\\2 & 1\end{smallmatrix}\right]$ & \makecell[l]{(5,6), (6,5), (6,6), (7,5), (7,6), (8,5), (8,6), (8,7) \\ (8,8), (8,9), (8,10), (9,5), (9,6), (9,10), (10,5)} \\ \cline{2-3}
    & $\left[\begin{smallmatrix}2 & 3\\-2 & 2\end{smallmatrix}\right]$ & \makecell[l]{(5,6), (6,5), (6,6), (7,5), (7,6), (8,5), (8,6), (8,7) \\ (8,8), (8,9), (8,10), (9,5), (9,6), (9,10), (10,5)} \\ \cline{2-3}
    & $\left[\begin{smallmatrix}5 & 0\\0 & 2\end{smallmatrix}\right]$ & \makecell[l]{(5,6), (6,5), (6,6), (7,5), (7,6), (8,5), (8,6), (8,7) \\ (8,8), (8,9), (8,10), (9,5), (9,6), (9,10), (10,5)} \\ \cline{2-3}
    & $\left[\begin{smallmatrix}5 & 1\\0 & 2\end{smallmatrix}\right]$ & \makecell[l]{(5,6), (6,5), (6,6), (7,5), (7,6), (8,5), (8,6), (8,7) \\ (8,8), (8,9), (8,10), (9,5), (9,6), (9,10), (10,5)} \\ \cline{2-3}
    & $\left[\begin{smallmatrix}5 & 2\\0 & 2\end{smallmatrix}\right]$ & \makecell[l]{(5,6), (6,5), (6,6), (7,5), (7,6), (8,5), (8,6), (8,7) \\ (8,8), (8,9), (8,10), (9,5), (9,6), (9,10), (10,5)} \\ \hline
    \multirow{3}{*}{$11$} & $\left[\begin{smallmatrix}2 & -3\\3 & 1\end{smallmatrix}\right]$ & \makecell[l]{(6,6), (6,7), (7,6), (7,7), (8,6), (8,7), (8,10), (9,6) \\ (9,7), (9,8), (9,9), (9,10), (9,11), (10,6), (10,11)} \\
    & $\left[\begin{smallmatrix}3 & -4\\2 & 1\end{smallmatrix}\right]$ & \makecell[l]{(6,6), (6,7), (7,6), (7,7), (8,6), (8,7), (8,9), (9,6) \\ (9,7), (9,8), (9,9), (9,10), (9,11), (10,6), (10,11)} \\ \cline{2-3}
    & $\left[\begin{smallmatrix}5 & -1\\1 & 2\end{smallmatrix}\right]$ & \makecell[l]{(6,6), (6,7), (7,6), (7,7), (8,6), (8,7), (8,10), (9,6) \\ (9,7), (9,8), (9,9), (9,10), (9,11), (10,6), (10,11)} \\ \hline
    \multirow{11}{*}{$12$} & $\left[\begin{smallmatrix}2 & 3\\-2 & 3\end{smallmatrix}\right]$ & \makecell[l]{(6,7), (7,7), (7,8), (8,6), (8,7), (9,6), (9,8), (9,11) \\ (10,6), (10,7), (10,9), (10,12), (11,6), (11,12), (12,6)} \\
    & $\left[\begin{smallmatrix}2 & 4\\-2 & 2\end{smallmatrix}\right]$ & \makecell[l]{(6,7), (7,6), (7,7), (8,6), (8,7), (9,6), (9,7), (9,9) \\ (9,10), (10,6), (10,7), (10,9), (10,12), (11,12), (12,6)} \\ \cline{2-3}
    & $\left[\begin{smallmatrix}3 & -3\\2 & 2\end{smallmatrix}\right]$ & \makecell[l]{(6,7), (7,6), (7,7), (8,6), (8,7), (9,6), (9,7), (9,9) \\ (9,10), (10,6), (10,7), (10,9), (10,12), (11,12), (12,6)} \\ \cline{2-3}
    & $\left[\begin{smallmatrix}3 & -3\\3 & 1\end{smallmatrix}\right]$ & \makecell[l]{(6,7), (7,6), (7,7), (8,6), (8,7), (9,6), (9,7), (9,9) \\ (9,10), (10,6), (10,7), (10,9), (10,12), (11,12), (12,6)} \\ \cline{2-3}
    & $\left[\begin{smallmatrix}4 & -4\\2 & 1\end{smallmatrix}\right]$ & \makecell[l]{(6,7), (7,6), (7,7), (8,6), (8,7), (9,6), (9,7), (9,9) \\ (9,10), (10,6), (10,7), (10,9), (10,12), (11,12), (12,6)} \\ \cline{2-3}
    & $\left[\begin{smallmatrix}4 & 0\\0 & 3\end{smallmatrix}\right]$ & \makecell[l]{(6,7), (7,6), (7,7), (8,6), (8,7), (9,6), (9,7), (9,9) \\ (9,10), (10,6), (10,7), (10,9), (10,12), (11,12), (12,6)} \\ \cline{2-3}
    & $\left[\begin{smallmatrix}4 & 0\\1 & 3\end{smallmatrix}\right]$ & \makecell[l]{(6,7), (7,6), (7,7), (8,6), (8,7), (9,6), (9,7), (9,9) \\ (9,10), (10,6), (10,7), (10,9), (10,12), (11,12), (12,6)} \\ \cline{2-3}
    & $\left[\begin{smallmatrix}4 & 1\\0 & 3\end{smallmatrix}\right]$ & \makecell[l]{(6,7), (7,6), (7,7), (8,6), (8,7), (9,6), (9,7), (9,9) \\ (9,10), (10,6), (10,7), (10,9), (10,12), (11,12), (12,6)} \\ \cline{2-3}
    & $\left[\begin{smallmatrix}4 & 2\\0 & 3\end{smallmatrix}\right]$ & \makecell[l]{(6,7), (7,6), (7,7), (8,6), (8,7), (9,6), (9,7), (9,9) \\ (9,10), (10,6), (10,7), (10,9), (10,12), (11,12), (12,6)} \\ \cline{2-3}
    & $\left[\begin{smallmatrix}5 & -2\\1 & 2\end{smallmatrix}\right]$ & \makecell[l]{(6,7), (7,6), (7,7), (8,6), (8,7), (9,6), (9,7), (9,9) \\ (9,10), (10,6), (10,7), (10,9), (10,12), (11,12), (12,6)} \\ \cline{2-3}
    & $\left[\begin{smallmatrix}6 & 0\\1 & 2\end{smallmatrix}\right]$ & \makecell[l]{(6,7), (7,6), (7,7), (8,6), (8,7), (9,6), (9,7), (9,9) \\ (9,10), (10,6), (10,7), (10,9), (10,12), (11,12), (12,6)} \\ \hline
    \multirow{1}{*}{$16$} & $\left[\begin{smallmatrix}4 & 0\\0 & 4\end{smallmatrix}\right]$ & \makecell[l]{(8,9), (9,10), (10,8), (10,9), (10,10), (11,8), (11,9), (12,8) \\ (12,14), (13,8), (13,12), (13,15), (13,16), (14,8), (14,16)} \\ \hline
    \multirow{1}{*}{$25$} & $\left[\begin{smallmatrix}5 & 0\\0 & 5\end{smallmatrix}\right]$ & \makecell[l]{(13,15), (15,15), (16,13), (16,15), (17,13), (18,13), (19,13), (19,21) \\ (19,22), (20,13), (20,21), (20,25), (21,13), (21,25), (22,25)} \\ \hline
  \hline
\end{longtable}

\clearpage
\subsection{Ag(100)}

Table~\ref{tab:asop_ag} lists all symmetry-inequivalent supercell transformation
matrices $\mathbf{H}$ grouped by $A = |\det\mathbf{H}|$, together with the
specific $(N_\mathrm{Ag}, N_\mathrm{O})$ compositions explored under each.

\begin{longtable}{ccl}
  \caption{Supercell vectors $[u, v]$ and their explored compositions $(N_\mathrm{Ag}, N_\mathrm{O})$ in ASOP.\label{tab:asop_ag}} \\
  \hline
  $A$ & $[u, v]$ & \multicolumn{1}{c}{Explored $(N_\mathrm{Ag}, N_\mathrm{O})$} \\
  \hline
  \endfirsthead
  \hline
  $A$ & $[u, v]$ & \multicolumn{1}{c}{Explored $(N_\mathrm{Ag}, N_\mathrm{O})$} \\
  \hline
  \endhead
    \multirow{4}{*}{$4$} & $\left[\begin{smallmatrix}1 & 3\\-1 & 1\end{smallmatrix}\right]$ & \makecell[l]{(1,1), (1,2), (1,3), (1,4), (2,1), (2,2), (2,3), (2,4) \\ (3,1), (3,2), (3,3), (3,4), (4,1), (4,2), (4,3), (4,4)} \\ \cline{2-3}
    & $\left[\begin{smallmatrix}2 & -2\\1 & 1\end{smallmatrix}\right]$ & \makecell[l]{(1,1), (1,2), (1,3), (1,4), (2,1), (2,2), (2,3), (2,4) \\ (3,1), (3,2), (3,3), (3,4), (4,1), (4,2), (4,3), (4,4)} \\ \cline{2-3}
    & $\left[\begin{smallmatrix}2 & 0\\0 & 2\end{smallmatrix}\right]$ & \makecell[l]{(1,1), (1,2), (1,3), (1,4), (2,1), (2,2), (2,3), (2,4) \\ (3,1), (3,2), (3,3), (3,4), (4,1), (4,2), (4,3), (4,4)} \\ \cline{2-3}
    & $\left[\begin{smallmatrix}2 & 1\\0 & 2\end{smallmatrix}\right]$ & \makecell[l]{(1,1), (1,2), (1,3), (1,4), (2,1), (2,2), (2,3), (2,4) \\ (3,1), (3,2), (3,3), (3,4), (4,1), (4,2), (4,3), (4,4)} \\ \hline
    \multirow{2}{*}{$5$} & $\left[\begin{smallmatrix}1 & -2\\2 & 1\end{smallmatrix}\right]$ & \makecell[l]{(1,2), (1,3), (2,2), (2,3), (3,2), (3,3), (3,4), (4,2) \\ (4,3), (4,4), (5,1), (5,2), (5,3), (5,4), (5,5)} \\ \cline{2-3}
    & $\left[\begin{smallmatrix}2 & 3\\-1 & 1\end{smallmatrix}\right]$ & \makecell[l]{(1,3), (2,2), (2,3), (3,2), (3,3), (3,4), (4,2), (4,3) \\ (4,4), (4,5), (5,1), (5,2), (5,3), (5,4), (5,5)} \\ \hline
    \multirow{4}{*}{$6$} & $\left[\begin{smallmatrix}2 & -2\\2 & 1\end{smallmatrix}\right]$ & \makecell[l]{(2,3), (3,3), (4,2), (4,3), (4,4), (4,5), (5,3), (5,4) \\ (5,5), (5,6), (6,2), (6,3), (6,4), (6,5), (6,6)} \\ \cline{2-3}
    & $\left[\begin{smallmatrix}3 & 0\\0 & 2\end{smallmatrix}\right]$ & \makecell[l]{(1,3), (1,4), (2,3), (3,2), (3,3), (4,3), (4,4), (5,3) \\ (5,4), (5,5), (5,6), (6,2), (6,3), (6,4), (6,6)} \\ \cline{2-3}
    & $\left[\begin{smallmatrix}3 & 0\\1 & 2\end{smallmatrix}\right]$ & \makecell[l]{(2,3), (3,3), (3,4), (4,3), (4,4), (5,2), (5,3), (5,4) \\ (5,5), (5,6), (6,2), (6,3), (6,4), (6,5), (6,6)} \\ \cline{2-3}
    & $\left[\begin{smallmatrix}3 & 1\\0 & 2\end{smallmatrix}\right]$ & \makecell[l]{(2,2), (2,3), (3,3), (4,3), (4,4), (5,2), (5,3), (5,4) \\ (5,5), (5,6), (6,2), (6,3), (6,4), (6,5), (6,6)} \\ \hline
    \multirow{1}{*}{$7$} & $\left[\begin{smallmatrix}1 & 3\\-2 & 1\end{smallmatrix}\right]$ & \makecell[l]{(3,3), (4,3), (4,4), (5,3), (5,4), (5,5), (6,3), (6,4) \\ (6,5), (6,6), (7,2), (7,3), (7,4), (7,5), (7,7)} \\ \hline
    \multirow{7}{*}{$8$} & $\left[\begin{smallmatrix}1 & 3\\-2 & 2\end{smallmatrix}\right]$ & \makecell[l]{(4,4), (5,4), (5,5), (6,4), (6,5), (6,6), (7,3), (7,4) \\ (7,5), (7,6), (7,7), (8,3), (8,4), (8,5), (8,6)} \\ \cline{2-3}
    & $\left[\begin{smallmatrix}2 & -3\\2 & 1\end{smallmatrix}\right]$ & \makecell[l]{(4,4), (5,4), (5,5), (6,4), (6,5), (6,6), (7,3), (7,4) \\ (7,5), (7,6), (7,7), (8,3), (8,4), (8,5), (8,6)} \\ \cline{2-3}
    & $\left[\begin{smallmatrix}2 & -2\\2 & 2\end{smallmatrix}\right]$ & \makecell[l]{(4,4), (5,4), (5,5), (6,4), (6,5), (6,6), (7,3), (7,4) \\ (7,5), (7,6), (7,7), (8,3), (8,4), (8,5), (8,6)} \\ \cline{2-3}
    & $\left[\begin{smallmatrix}4 & 0\\0 & 2\end{smallmatrix}\right]$ & \makecell[l]{(4,4), (5,4), (5,5), (6,4), (6,5), (6,6), (7,3), (7,4) \\ (7,5), (7,6), (7,7), (8,3), (8,4), (8,5), (8,6)} \\ \cline{2-3}
    & $\left[\begin{smallmatrix}4 & 0\\1 & 2\end{smallmatrix}\right]$ & \makecell[l]{(4,4), (5,4), (5,5), (6,4), (6,5), (6,6), (7,3), (7,4) \\ (7,5), (7,6), (7,7), (8,3), (8,4), (8,5), (8,6)} \\ \cline{2-3}
    & $\left[\begin{smallmatrix}4 & 1\\0 & 2\end{smallmatrix}\right]$ & \makecell[l]{(4,4), (5,4), (5,5), (6,4), (6,5), (6,6), (7,3), (7,4) \\ (7,5), (7,6), (7,7), (8,3), (8,4), (8,5), (8,6)} \\ \cline{2-3}
    & $\left[\begin{smallmatrix}4 & 2\\0 & 2\end{smallmatrix}\right]$ & \makecell[l]{(4,4), (5,4), (5,5), (6,4), (6,5), (6,6), (7,3), (7,4) \\ (7,5), (7,6), (7,7), (8,3), (8,4), (8,5), (8,6)} \\ \hline
    \multirow{4}{*}{$9$} & $\left[\begin{smallmatrix}1 & 4\\-2 & 1\end{smallmatrix}\right]$ & \makecell[l]{(4,4), (5,5), (6,4), (6,5), (7,4), (7,5), (7,6), (7,7) \\ (8,4), (8,5), (8,6), (8,7), (8,8), (9,4), (9,5)} \\ \cline{2-3}
    & $\left[\begin{smallmatrix}3 & -3\\2 & 1\end{smallmatrix}\right]$ & \makecell[l]{(5,5), (6,4), (6,5), (7,4), (7,5), (7,6), (7,7), (8,4) \\ (8,5), (8,6), (8,7), (8,8), (9,3), (9,4), (9,5)} \\ \cline{2-3}
    & $\left[\begin{smallmatrix}3 & 0\\0 & 3\end{smallmatrix}\right]$ & \makecell[l]{(5,5), (6,4), (6,5), (7,4), (7,5), (7,6), (7,7), (8,4) \\ (8,5), (8,6), (8,7), (8,8), (9,3), (9,4), (9,5)} \\ \cline{2-3}
    & $\left[\begin{smallmatrix}3 & 1\\0 & 3\end{smallmatrix}\right]$ & \makecell[l]{(5,5), (6,4), (6,5), (7,4), (7,5), (7,6), (7,7), (8,4) \\ (8,5), (8,6), (8,7), (8,8), (9,3), (9,4), (9,5)} \\ \hline
    \multirow{6}{*}{$10$} & $\left[\begin{smallmatrix}1 & -3\\3 & 1\end{smallmatrix}\right]$ & \makecell[l]{(6,5), (6,6), (7,5), (7,6), (8,5), (8,6), (8,7), (8,8) \\ (9,5), (9,6), (9,7), (9,8), (9,9), (10,5), (10,6)} \\ \cline{2-3}
    & $\left[\begin{smallmatrix}2 & -4\\2 & 1\end{smallmatrix}\right]$ & \makecell[l]{(6,5), (6,6), (7,5), (7,6), (8,5), (8,6), (8,7), (8,8) \\ (9,4), (9,5), (9,6), (9,7), (9,8), (10,5), (10,6)} \\ \cline{2-3}
    & $\left[\begin{smallmatrix}2 & 3\\-2 & 2\end{smallmatrix}\right]$ & \makecell[l]{(5,5), (6,5), (6,6), (7,5), (7,6), (8,5), (8,6), (8,7) \\ (8,8), (9,5), (9,6), (9,7), (9,8), (10,5), (10,6)} \\ \cline{2-3}
    & $\left[\begin{smallmatrix}5 & 0\\0 & 2\end{smallmatrix}\right]$ & \makecell[l]{(5,5), (6,5), (6,6), (7,5), (7,6), (8,5), (8,6), (8,7) \\ (8,8), (9,5), (9,6), (9,7), (9,8), (10,5), (10,6)} \\ \cline{2-3}
    & $\left[\begin{smallmatrix}5 & 1\\0 & 2\end{smallmatrix}\right]$ & \makecell[l]{(5,5), (6,5), (6,6), (7,5), (7,6), (8,5), (8,6), (8,7) \\ (8,8), (9,5), (9,6), (9,7), (9,8), (10,5), (10,6)} \\ \cline{2-3}
    & $\left[\begin{smallmatrix}5 & 2\\0 & 2\end{smallmatrix}\right]$ & \makecell[l]{(5,5), (6,5), (6,6), (7,5), (7,6), (8,5), (8,6), (8,7) \\ (8,8), (9,5), (9,6), (9,7), (9,8), (10,5), (10,6)} \\ \hline
    \multirow{3}{*}{$11$} & $\left[\begin{smallmatrix}2 & -3\\3 & 1\end{smallmatrix}\right]$ & \makecell[l]{(7,6), (8,5), (8,6), (8,7), (8,8), (9,6), (9,7), (9,8) \\ (9,9), (10,5), (10,6), (10,7), (10,8), (10,9), (11,5)} \\ \cline{2-3}
    & $\left[\begin{smallmatrix}3 & -4\\2 & 1\end{smallmatrix}\right]$ & \makecell[l]{(7,5), (7,6), (7,7), (8,6), (8,7), (9,6), (9,7), (9,8) \\ (9,9), (10,5), (10,6), (10,7), (10,8), (10,9), (11,5)} \\ \cline{2-3}
    & $\left[\begin{smallmatrix}5 & -1\\1 & 2\end{smallmatrix}\right]$ & \makecell[l]{(7,6), (7,7), (8,6), (8,7), (9,6), (9,7), (9,8), (9,9) \\ (10,5), (10,6), (10,7), (10,8), (10,9), (11,5), (11,7)} \\ \hline
    \multirow{11}{*}{$12$} & $\left[\begin{smallmatrix}2 & 3\\-2 & 3\end{smallmatrix}\right]$ & \makecell[l]{(7,6), (7,7), (8,7), (9,6), (9,7), (9,8), (9,9), (10,7) \\ (10,8), (10,9), (11,6), (11,7), (11,8), (11,9), (12,6)} \\ \cline{2-3}
    & $\left[\begin{smallmatrix}2 & 4\\-2 & 2\end{smallmatrix}\right]$ & \makecell[l]{(7,7), (8,7), (9,6), (9,7), (9,8), (9,9), (10,7), (10,8) \\ (10,9), (10,10), (11,6), (11,7), (11,8), (11,10), (12,6)} \\ \cline{2-3}
    & $\left[\begin{smallmatrix}3 & -3\\2 & 2\end{smallmatrix}\right]$ & \makecell[l]{(7,6), (7,7), (8,7), (9,6), (9,7), (9,8), (9,9), (10,7) \\ (10,8), (10,9), (10,10), (11,6), (11,7), (11,8), (12,6)} \\ \cline{2-3}
    & $\left[\begin{smallmatrix}3 & -3\\3 & 1\end{smallmatrix}\right]$ & \makecell[l]{(7,6), (7,7), (8,7), (9,6), (9,7), (9,8), (9,9), (10,7) \\ (10,8), (10,9), (10,10), (11,6), (11,7), (11,8), (12,6)} \\ \cline{2-3}
    & $\left[\begin{smallmatrix}4 & -4\\2 & 1\end{smallmatrix}\right]$ & \makecell[l]{(7,6), (7,7), (8,7), (9,6), (9,7), (9,8), (9,9), (10,7) \\ (10,8), (10,9), (10,10), (11,6), (11,7), (11,8), (12,6)} \\ \cline{2-3}
    & $\left[\begin{smallmatrix}4 & 0\\0 & 3\end{smallmatrix}\right]$ & \makecell[l]{(7,6), (7,7), (8,7), (9,6), (9,7), (9,8), (9,9), (10,7) \\ (10,8), (10,9), (10,10), (11,6), (11,7), (11,8), (12,6)} \\ \cline{2-3}
    & $\left[\begin{smallmatrix}4 & 0\\1 & 3\end{smallmatrix}\right]$ & \makecell[l]{(7,6), (7,7), (8,7), (9,6), (9,7), (9,8), (9,9), (10,7) \\ (10,8), (10,9), (10,10), (11,6), (11,7), (11,8), (12,6)} \\ \cline{2-3}
    & $\left[\begin{smallmatrix}4 & 1\\0 & 3\end{smallmatrix}\right]$ & \makecell[l]{(7,6), (7,7), (8,7), (9,6), (9,7), (9,8), (9,9), (10,7) \\ (10,8), (10,9), (10,10), (11,6), (11,7), (11,8), (12,6)} \\ \cline{2-3}
    & $\left[\begin{smallmatrix}4 & 2\\0 & 3\end{smallmatrix}\right]$ & \makecell[l]{(7,6), (7,7), (8,7), (9,6), (9,7), (9,8), (9,9), (10,7) \\ (10,8), (10,9), (10,10), (11,6), (11,7), (11,8), (12,6)} \\ \cline{2-3}
    & $\left[\begin{smallmatrix}5 & -2\\1 & 2\end{smallmatrix}\right]$ & \makecell[l]{(7,6), (7,7), (8,7), (9,6), (9,7), (9,8), (9,9), (10,7) \\ (10,8), (10,9), (10,10), (11,6), (11,7), (11,8), (12,6)} \\ \cline{2-3}
    & $\left[\begin{smallmatrix}6 & 0\\1 & 2\end{smallmatrix}\right]$ & \makecell[l]{(7,6), (7,7), (8,7), (9,6), (9,7), (9,8), (9,9), (10,7) \\ (10,8), (10,9), (10,10), (11,6), (11,7), (11,8), (12,6)} \\ \hline
    \multirow{1}{*}{$16$} & $\left[\begin{smallmatrix}4 & 0\\0 & 4\end{smallmatrix}\right]$ & \makecell[l]{(10,9), (12,8), (12,9), (12,10), (12,11), (13,9), (13,10), (13,11) \\ (13,12), (13,13), (14,9), (14,10), (14,11), (14,12), (14,13)} \\ \hline
    \multirow{1}{*}{$25$} & $\left[\begin{smallmatrix}5 & 0\\0 & 5\end{smallmatrix}\right]$ & \makecell[l]{(19,13), (19,15), (19,16), (20,15), (20,18), (20,19), (20,20), (21,15) \\ (21,16), (21,17), (21,18), (21,19), (22,15), (22,16), (22,17)} \\ \hline
  \hline
\end{longtable}

\clearpage
\subsection{Ag(100) with ethylene}

Table~\ref{tab:asop_ag_reaction} lists all symmetry-inequivalent supercell transformation
matrices $\mathbf{H}$ grouped by $A = |\det\mathbf{H}|$, together with the
specific $(N_\mathrm{Ag}, N_\mathrm{O}, N_\mathrm{C_2H_4})$ compositions explored under each.

\begin{longtable}{ccl}
  \caption{Supercell vectors $[u, v]$ and their explored compositions $(N_\mathrm{Ag}, N_\mathrm{O}, N_\mathrm{C_2H_4})$ in ASOP for Ag(100) with ethylene.\label{tab:asop_ag_reaction}} \\
  \hline
  $A$ & $[u, v]$ & \multicolumn{1}{c}{Explored $(N_\mathrm{Ag}, N_\mathrm{O}, N_\mathrm{C_2H_4})$} \\
  \hline
  \endfirsthead
  \hline
  $A$ & $[u, v]$ & \multicolumn{1}{c}{Explored $(N_\mathrm{Ag}, N_\mathrm{O}, N_\mathrm{C_2H_4})$} \\
  \hline
  \endhead
    \multirow{4}{*}{$4$} & $\left[\begin{smallmatrix}1 & 3\\-1 & 1\end{smallmatrix}\right]$ & \makecell[l]{(1,1,1), (1,2,1), (1,3,1), (1,4,1), (2,1,1), (2,2,1), (2,3,1), (2,4,1) \\ (3,1,1), (3,2,1), (3,3,1), (3,4,1), (4,1,1), (4,2,1), (4,3,1), (4,4,1)} \\ \cline{2-3}
    & $\left[\begin{smallmatrix}2 & -2\\1 & 1\end{smallmatrix}\right]$ & \makecell[l]{(1,1,1), (1,2,1), (1,3,1), (1,4,1), (2,1,1), (2,2,1), (2,3,1), (2,4,1) \\ (3,1,1), (3,2,1), (3,3,1), (3,4,1), (4,1,1), (4,2,1), (4,3,1), (4,4,1)} \\ \cline{2-3}
    & $\left[\begin{smallmatrix}2 & 0\\0 & 2\end{smallmatrix}\right]$ & \makecell[l]{(1,1,1), (1,2,1), (1,3,1), (1,4,1), (2,1,1), (2,2,1), (2,3,1), (2,4,1) \\ (3,1,1), (3,2,1), (3,3,1), (3,4,1), (4,1,1), (4,2,1), (4,3,1), (4,4,1)} \\ \cline{2-3}
    & $\left[\begin{smallmatrix}2 & 1\\0 & 2\end{smallmatrix}\right]$ & \makecell[l]{(1,1,1), (1,2,1), (1,3,1), (1,4,1), (2,1,1), (2,2,1), (2,3,1), (2,4,1) \\ (3,1,1), (3,2,1), (3,3,1), (3,4,1), (4,1,1), (4,2,1), (4,3,1), (4,4,1)} \\ \hline
    \multirow{2}{*}{$5$} & $\left[\begin{smallmatrix}1 & -2\\2 & 1\end{smallmatrix}\right]$ & \makecell[l]{(1,3,1), (1,4,1), (2,3,1), (2,4,1), (3,2,1), (3,3,1), (3,4,1), (4,2,1) \\ (4,3,1), (4,4,1), (5,1,1), (5,2,1), (5,3,1), (5,4,1), (5,5,1)} \\ \cline{2-3}
    & $\left[\begin{smallmatrix}2 & 3\\-1 & 1\end{smallmatrix}\right]$ & \makecell[l]{(1,3,1), (3,1,1), (3,2,1), (3,3,1), (3,4,1), (4,1,1), (4,2,1), (4,3,1) \\ (4,4,1), (4,5,1), (5,1,1), (5,2,1), (5,3,1), (5,4,1), (5,5,1)} \\ \hline
    \multirow{4}{*}{$6$} & $\left[\begin{smallmatrix}2 & -2\\2 & 1\end{smallmatrix}\right]$ & \makecell[l]{(3,3,1), (4,2,1), (4,3,1), (4,4,1), (4,5,1), (5,2,1), (5,3,1), (5,4,1) \\ (5,5,1), (5,6,1), (6,1,1), (6,2,1), (6,3,1), (6,4,1), (6,6,1)} \\ \cline{2-3}
    & $\left[\begin{smallmatrix}3 & 0\\0 & 2\end{smallmatrix}\right]$ & \makecell[l]{(3,2,1), (3,3,1), (3,4,1), (4,2,1), (4,3,1), (4,4,1), (5,2,1), (5,3,1) \\ (5,4,1), (5,5,1), (5,6,1), (6,1,1), (6,2,1), (6,3,1), (6,4,1)} \\ \cline{2-3}
    & $\left[\begin{smallmatrix}3 & 0\\1 & 2\end{smallmatrix}\right]$ & \makecell[l]{(3,2,1), (3,3,1), (3,4,1), (4,2,1), (4,3,1), (4,4,1), (5,2,1), (5,3,1) \\ (5,4,1), (5,5,1), (5,6,1), (6,1,1), (6,2,1), (6,3,1), (6,4,1)} \\ \cline{2-3}
    & $\left[\begin{smallmatrix}3 & 1\\0 & 2\end{smallmatrix}\right]$ & \makecell[l]{(3,2,1), (3,3,1), (3,4,1), (4,2,1), (4,3,1), (4,4,1), (5,2,1), (5,3,1) \\ (5,4,1), (5,5,1), (5,6,1), (6,1,1), (6,2,1), (6,3,1), (6,4,1)} \\ \hline
    \multirow{1}{*}{$7$} & $\left[\begin{smallmatrix}1 & 3\\-2 & 1\end{smallmatrix}\right]$ & \makecell[l]{(2,5,2), (4,3,2), (4,4,2), (5,3,2), (5,4,2), (5,5,2), (5,6,2), (6,3,2) \\ (6,4,2), (6,5,2), (6,6,2), (7,3,2), (7,4,2), (7,5,2), (7,6,2)} \\ \hline
    \multirow{7}{*}{$8$} & $\left[\begin{smallmatrix}1 & 3\\-2 & 2\end{smallmatrix}\right]$ & \makecell[l]{(5,4,2), (5,5,2), (6,3,2), (6,4,2), (6,5,2), (6,6,2), (7,3,2), (7,4,2) \\ (7,5,2), (7,6,2), (8,2,2), (8,3,2), (8,4,2), (8,5,2), (8,6,2)} \\ \cline{2-3}
    & $\left[\begin{smallmatrix}2 & -3\\2 & 1\end{smallmatrix}\right]$ & \makecell[l]{(5,4,2), (5,5,2), (6,3,2), (6,4,2), (6,5,2), (6,6,2), (7,3,2), (7,4,2) \\ (7,5,2), (7,6,2), (8,2,2), (8,3,2), (8,4,2), (8,5,2), (8,6,2)} \\ \cline{2-3}
    & $\left[\begin{smallmatrix}2 & -2\\2 & 2\end{smallmatrix}\right]$ & \makecell[l]{(5,4,2), (5,5,2), (6,3,2), (6,4,2), (6,5,2), (6,6,2), (7,3,2), (7,4,2) \\ (7,5,2), (7,6,2), (8,2,2), (8,3,2), (8,4,2), (8,5,2), (8,6,2)} \\ \cline{2-3}
    & $\left[\begin{smallmatrix}4 & 0\\0 & 2\end{smallmatrix}\right]$ & \makecell[l]{(5,4,2), (5,5,2), (6,3,2), (6,4,2), (6,5,2), (6,6,2), (7,3,2), (7,4,2) \\ (7,5,2), (7,6,2), (8,2,2), (8,3,2), (8,4,2), (8,5,2), (8,6,2)} \\ \cline{2-3}
    & $\left[\begin{smallmatrix}4 & 0\\1 & 2\end{smallmatrix}\right]$ & \makecell[l]{(5,4,2), (5,5,2), (6,3,2), (6,4,2), (6,5,2), (6,6,2), (7,3,2), (7,4,2) \\ (7,5,2), (7,6,2), (8,2,2), (8,3,2), (8,4,2), (8,5,2), (8,6,2)} \\ \cline{2-3}
    & $\left[\begin{smallmatrix}4 & 1\\0 & 2\end{smallmatrix}\right]$ & \makecell[l]{(5,4,2), (5,5,2), (6,3,2), (6,4,2), (6,5,2), (6,6,2), (7,3,2), (7,4,2) \\ (7,5,2), (7,6,2), (8,2,2), (8,3,2), (8,4,2), (8,5,2), (8,6,2)} \\ \cline{2-3}
    & $\left[\begin{smallmatrix}4 & 2\\0 & 2\end{smallmatrix}\right]$ & \makecell[l]{(5,4,2), (5,5,2), (6,3,2), (6,4,2), (6,5,2), (6,6,2), (7,3,2), (7,4,2) \\ (7,5,2), (7,6,2), (8,2,2), (8,3,2), (8,4,2), (8,5,2), (8,6,2)} \\ \hline
    \multirow{4}{*}{$9$} & $\left[\begin{smallmatrix}1 & 4\\-2 & 1\end{smallmatrix}\right]$ & \makecell[l]{(6,4,2), (6,5,2), (6,6,2), (7,4,2), (7,5,2), (7,6,2), (7,7,2), (8,4,2) \\ (8,5,2), (8,6,2), (8,7,2), (9,3,2), (9,4,2), (9,6,2), (9,7,2)} \\ \cline{2-3}
    & $\left[\begin{smallmatrix}3 & -3\\2 & 1\end{smallmatrix}\right]$ & \makecell[l]{(5,6,2), (6,5,2), (6,6,2), (7,4,2), (7,5,2), (7,6,2), (7,7,2), (8,4,2) \\ (8,5,2), (8,6,2), (9,3,2), (9,4,2), (9,5,2), (9,6,2), (9,7,2)} \\ \cline{2-3}
    & $\left[\begin{smallmatrix}3 & 0\\0 & 3\end{smallmatrix}\right]$ & \makecell[l]{(6,5,2), (6,6,2), (7,4,2), (7,5,2), (7,6,2), (7,7,2), (8,4,2), (8,5,2) \\ (8,6,2), (8,7,2), (9,3,2), (9,4,2), (9,5,2), (9,6,2), (9,7,2)} \\ \cline{2-3}
    & $\left[\begin{smallmatrix}3 & 1\\0 & 3\end{smallmatrix}\right]$ & \makecell[l]{(6,5,2), (6,6,2), (7,4,2), (7,5,2), (7,6,2), (7,7,2), (8,4,2), (8,5,2) \\ (8,6,2), (8,7,2), (9,3,2), (9,4,2), (9,5,2), (9,6,2), (9,7,2)} \\ \hline
    \multirow{6}{*}{$10$} & $\left[\begin{smallmatrix}1 & -3\\3 & 1\end{smallmatrix}\right]$ & \makecell[l]{(6,6,3), (7,5,3), (7,6,3), (8,4,3), (8,5,3), (8,6,3), (8,7,3), (9,4,3) \\ (9,5,3), (9,6,3), (9,7,2), (9,7,3), (10,4,3), (10,7,2), (10,8,3)} \\ \cline{2-3}
    & $\left[\begin{smallmatrix}2 & -4\\2 & 1\end{smallmatrix}\right]$ & \makecell[l]{(6,6,3), (7,5,3), (7,6,3), (8,5,3), (8,6,2), (8,6,3), (8,7,3), (8,8,2) \\ (9,4,3), (9,5,3), (9,6,2), (9,6,3), (9,7,3), (10,4,3), (10,5,3)} \\ \cline{2-3}
    & $\left[\begin{smallmatrix}2 & 3\\-2 & 2\end{smallmatrix}\right]$ & \makecell[l]{(6,6,3), (7,5,3), (7,6,3), (8,5,3), (8,6,2), (8,6,3), (8,7,3), (8,8,2) \\ (9,4,3), (9,5,3), (9,6,2), (9,6,3), (9,7,3), (10,3,3), (10,4,3)} \\ \cline{2-3}
    & $\left[\begin{smallmatrix}5 & 0\\0 & 2\end{smallmatrix}\right]$ & \makecell[l]{(6,6,3), (7,5,3), (7,6,3), (8,5,3), (8,6,2), (8,6,3), (8,7,3), (8,8,2) \\ (9,4,3), (9,5,3), (9,6,2), (9,6,3), (9,7,3), (10,3,3), (10,4,3)} \\ \cline{2-3}
    & $\left[\begin{smallmatrix}5 & 1\\0 & 2\end{smallmatrix}\right]$ & \makecell[l]{(6,6,3), (7,5,3), (7,6,3), (8,5,3), (8,6,2), (8,6,3), (8,7,3), (8,8,2) \\ (9,4,3), (9,5,3), (9,6,2), (9,6,3), (9,7,3), (10,3,3), (10,4,3)} \\ \cline{2-3}
    & $\left[\begin{smallmatrix}5 & 2\\0 & 2\end{smallmatrix}\right]$ & \makecell[l]{(6,6,3), (7,5,3), (7,6,3), (8,5,3), (8,6,2), (8,6,3), (8,7,3), (8,8,2) \\ (9,4,3), (9,5,3), (9,6,2), (9,6,3), (9,7,3), (10,3,3), (10,4,3)} \\ \hline
    \multirow{3}{*}{$11$} & $\left[\begin{smallmatrix}2 & -3\\3 & 1\end{smallmatrix}\right]$ & \makecell[l]{(7,6,3), (7,7,3), (8,5,3), (8,6,3), (8,7,3), (9,6,3), (9,7,3), (9,8,3) \\ (10,4,3), (10,5,3), (10,6,3), (10,7,3), (10,8,3), (11,4,3), (11,8,3)} \\ \cline{2-3}
    & $\left[\begin{smallmatrix}3 & -4\\2 & 1\end{smallmatrix}\right]$ & \makecell[l]{(7,6,3), (7,7,3), (8,5,3), (8,6,3), (8,7,3), (8,8,3), (9,6,3), (9,7,3) \\ (9,8,3), (10,5,3), (10,6,3), (10,7,3), (10,8,3), (11,4,3), (11,5,3)} \\ \cline{2-3}
    & $\left[\begin{smallmatrix}5 & -1\\1 & 2\end{smallmatrix}\right]$ & \makecell[l]{(7,6,3), (7,7,3), (8,5,3), (8,6,3), (8,7,3), (9,5,3), (9,6,3), (9,7,3) \\ (9,8,3), (10,5,3), (10,6,3), (10,7,3), (10,8,3), (11,4,3), (11,5,3)} \\ \hline
    \multirow{11}{*}{$12$} & $\left[\begin{smallmatrix}2 & 3\\-2 & 3\end{smallmatrix}\right]$ & \makecell[l]{(8,6,3), (8,7,3), (9,6,3), (9,7,3), (9,8,3), (10,7,3), (10,8,3), (10,9,3) \\ (11,5,3), (11,6,3), (11,7,3), (11,8,3), (11,9,3), (12,6,3), (12,9,3)} \\ \cline{2-3}
    & $\left[\begin{smallmatrix}2 & 4\\-2 & 2\end{smallmatrix}\right]$ & \makecell[l]{(7,7,3), (8,7,3), (9,6,3), (9,7,3), (9,8,3), (9,9,3), (10,6,3), (10,7,3) \\ (10,8,3), (10,9,3), (11,6,3), (11,8,3), (11,9,3), (12,6,3), (12,9,3)} \\ \cline{2-3}
    & $\left[\begin{smallmatrix}3 & -3\\2 & 2\end{smallmatrix}\right]$ & \makecell[l]{(7,7,3), (8,7,3), (9,6,3), (9,7,3), (9,8,3), (9,9,3), (10,6,3), (10,7,3) \\ (10,8,3), (10,9,3), (11,6,3), (11,8,3), (11,9,3), (12,6,3), (12,9,3)} \\ \cline{2-3}
    & $\left[\begin{smallmatrix}3 & -3\\3 & 1\end{smallmatrix}\right]$ & \makecell[l]{(7,7,3), (8,7,3), (9,6,3), (9,7,3), (9,8,3), (9,9,3), (10,6,3), (10,7,3) \\ (10,8,3), (10,9,3), (11,6,3), (11,8,3), (11,9,3), (12,6,3), (12,9,3)} \\ \cline{2-3}
    & $\left[\begin{smallmatrix}4 & -4\\2 & 1\end{smallmatrix}\right]$ & \makecell[l]{(7,7,3), (8,7,3), (9,6,3), (9,7,3), (9,8,3), (9,9,3), (10,6,3), (10,7,3) \\ (10,8,3), (10,9,3), (11,6,3), (11,8,3), (11,9,3), (12,6,3), (12,9,3)} \\ \cline{2-3}
    & $\left[\begin{smallmatrix}4 & 0\\0 & 3\end{smallmatrix}\right]$ & \makecell[l]{(7,7,3), (8,7,3), (9,6,3), (9,7,3), (9,8,3), (9,9,3), (10,6,3), (10,7,3) \\ (10,8,3), (10,9,3), (11,6,3), (11,8,3), (11,9,3), (12,6,3), (12,9,3)} \\ \cline{2-3}
    & $\left[\begin{smallmatrix}4 & 0\\1 & 3\end{smallmatrix}\right]$ & \makecell[l]{(7,7,3), (8,7,3), (9,6,3), (9,7,3), (9,8,3), (9,9,3), (10,6,3), (10,7,3) \\ (10,8,3), (10,9,3), (11,6,3), (11,8,3), (11,9,3), (12,6,3), (12,9,3)} \\ \cline{2-3}
    & $\left[\begin{smallmatrix}4 & 1\\0 & 3\end{smallmatrix}\right]$ & \makecell[l]{(7,7,3), (8,7,3), (9,6,3), (9,7,3), (9,8,3), (9,9,3), (10,6,3), (10,7,3) \\ (10,8,3), (10,9,3), (11,6,3), (11,8,3), (11,9,3), (12,6,3), (12,9,3)} \\ \cline{2-3}
    & $\left[\begin{smallmatrix}4 & 2\\0 & 3\end{smallmatrix}\right]$ & \makecell[l]{(7,7,3), (8,7,3), (9,6,3), (9,7,3), (9,8,3), (9,9,3), (10,6,3), (10,7,3) \\ (10,8,3), (10,9,3), (11,6,3), (11,8,3), (11,9,3), (12,6,3), (12,9,3)} \\ \cline{2-3}
    & $\left[\begin{smallmatrix}5 & -2\\1 & 2\end{smallmatrix}\right]$ & \makecell[l]{(7,7,3), (8,7,3), (9,6,3), (9,7,3), (9,8,3), (9,9,3), (10,6,3), (10,7,3) \\ (10,8,3), (10,9,3), (11,6,3), (11,8,3), (11,9,3), (12,6,3), (12,9,3)} \\ \cline{2-3}
    & $\left[\begin{smallmatrix}6 & 0\\1 & 2\end{smallmatrix}\right]$ & \makecell[l]{(7,7,3), (8,7,3), (9,6,3), (9,7,3), (9,8,3), (9,9,3), (10,6,3), (10,7,3) \\ (10,8,3), (10,9,3), (11,6,3), (11,8,3), (11,9,3), (12,6,3), (12,9,3)} \\ \hline
    \multirow{1}{*}{$16$} & $\left[\begin{smallmatrix}4 & 0\\0 & 4\end{smallmatrix}\right]$ & \makecell[l]{(10,9,5), (11,8,5), (11,9,5), (12,8,5), (12,9,5), (12,10,5), (13,9,5), (13,10,5) \\ (13,11,4), (13,11,5), (14,9,5), (14,10,5), (14,11,4), (14,11,5), (16,12,4)} \\ \hline
    \multirow{1}{*}{$25$} & $\left[\begin{smallmatrix}5 & 0\\0 & 5\end{smallmatrix}\right]$ & \makecell[l]{(19,15,7), (20,14,7), (20,15,6), (20,15,7), (20,18,7), (20,19,7), (20,19,8), (21,15,6) \\ (21,16,6), (21,18,7), (21,19,7), (21,19,8), (23,11,8), (24,19,6), (25,19,6)} \\ \hline
  \hline
\end{longtable}

\end{document}